\documentclass[final,3p,12pt]{elsarticle}

\usepackage{amssymb}
\usepackage{amsmath}
\providecommand*{\diff}{\,\text{d}}
\providecommand*{\ped}[1]{\ensuremath{_\mathrm{#1}}}    % text subscript
\providecommand*{\unit}[1]{\ensuremath{\mathrm{\,#1}}}

\newcommand{\revonecolor}{black}
\newcommand{\revtwocolor}{black}
\newcommand{\revone}[1]{\textcolor{\revonecolor}{#1}}
\newcommand{\revtwo}[1]{\textcolor{\revtwocolor}{#1}}

\newlength{\AppCImgWidth}
\usepackage[dvipsnames]{xcolor}

\usepackage{subcaption}

\usepackage{floatflt}

\usepackage{bm}

\usepackage{float}

\usepackage{layouts}

\usepackage[hyphens]{url}

\usepackage{upgreek}

\usepackage{lmodern}

\usepackage{url}

\journal{International Journal of Heat and Mass Transfer}

\usepackage{cleveref}
\crefname{figure}{Figure}{Figures}
\crefname{equation}{Equation}{Equations}
\crefname{table}{Table}{Tables}
\crefname{section}{Section}{Sections}
\crefname{appendix}{}{}

\begin{document}

    \begin{frontmatter}
        
        
        
        \title{\revone{A novel simulation approach for concentration-driven evaporation in capillaries}}
        
        
        \author[label1]{Phil Namesnik}
        \author[label1]{Alexander Eifert}
        \author[label1]{Anja Lippert\corref{cor1}}\ead{Anja.Lippert@bosch.com}\cortext[cor1]{Anja.Lippert@bosch.com}
        \author[label1]{Louis Mett}
        \author[label2]{Uwe Janoske}\ead{janoske@uni-wuppertal.de}
        
        \affiliation[label1]{organization={Robert Bosch GmbH},
                     addressline={Robert-Bosch-Campus 1},
                     city={Renningen},
                     postcode={71272},
                     country={Germany}}
        \affiliation[label2]{organization={School of Mechanical and Safety Engineering, University Wuppertal},
                     addressline={Gaußstraße 20},
                     city={Wuppertal},
                     postcode={42119},
                     country={Germany}}
        
        % \author{\fnm{Phil} \sur{Namesnik}\fnref{label2}}
        % \affiliation{organization={Robert Bosch GmbH},
        %              addressline={Robert-Bosch-Campus 1},
        %              city={Renningen},
        %              postcode={71272},
        %              country={Germany}}

        % \author{\fnm{Alexander} \sur{Eifert}\fnref{label2}}
        
        % \author{\fnm{Anja} \sur{Lippert}\corref{cor1}\fnref{label2}}\ead{Anja.Lippert@bosch.com}
        % \cortext[cor1]{Anja.Lippert@bosch.com}
        
        % \author{\fnm{Tobias} \sur{Tolle}\fnref{label2}}

        % \author{\fnm{Louis} \sur{Mett}\fnref{label2}}
        
        % \author{\fnm{Uwe} \sur{Janoske} \fnref{label3}}
        % \ead{janoske@uni-wuppertal.de}
        % \fntext[label3]{ }
        % \affiliation{organization={School of Mechanical and Safety Engineering, University Wuppertal},
        %              addressline={Gaußstraße 20},
        %              city={Wuppertal},
        %              postcode={42119},
        %              country={Germany}}

        \begin{abstract}
            Long liquid retention times in industrial gaps, due to capillary effects, significantly affect product lifetime by facilitating corrosion on solid surfaces. Concentration-driven evaporation plays a major role in mitigating this corrosion. Accurate evaporation rate predictions are crucial for improved product design. However, simulating capillary-driven flows with evaporation in complex geometries is challenging, requiring consideration of surface tension, wetting, and phase-change effects. Traditional approaches, such as the Volume-of-Fluid method, are prone to curvature calculation errors and have long simulation times due to strict time step limitations.

This study introduces a novel semi-transient simulation approach for fast evaporation rate prediction in arbitrarily shaped cavities. The approach involves a unidirectional coupling circuit, simulating the fluid surface in Surface Evolver and combining it with a vapor-in-gas diffusion simulation in OpenFOAM. The approach assumes that the evaporation rate is calculated solely based on the conditions at a given liquid filling level, without considering the evaporation history. This allows for highly parallelized simulations, achieving simulation runtimes in the order of 10 minutes to cover up to 150 hours of physical time.

Numerical investigations are conducted for water evaporation in air at a temperature of 23 °C and a relative humidity of 17 \%, for round and polygonal-shaped capillaries with inner diameters ranging from 1 mm to 13 mm. The results are validated using experimental data and show strong agreement. Simulations are also performed for complex industrial relevant gaps, demonstrating the applicability of the approach to a wide range of crevice geometries.
        \end{abstract}
        
        %     \includegraphics[width=\linewidth]{grabs.pdf}
        
        % \begin{highlights}
        %     \item Novel simulation approach to predict concentration-driven evaporation rates from cavities governed by capillary effects
        %     \item Unidirectional coupling circuit utilizing Surface Evolver to simulate liquid surfaces combined with vapor-in-gas diffusion simulation in OpenFOAM
        %     \item Quasi-steady approach allows high degree of parallelization and fast computation
        %     \item Simulation results are validated against experimental data and show strong agreement
        % \end{highlights}
        
    
    \end{frontmatter}
    
    
    % Introduction - What is the usecase of this approach? Current research? Bridge to own approach
    \section{Introduction}
\label{sec:introduction}
    Long liquid retention times in industrial gaps enable corrosion at solid surfaces which can be mitigated by evaporation of the liquid. Volatile liquids evaporate at a free liquid-gas surface and lead to a mass loss. The rate $\dot{m}\ped{ev}$ at which a liquid evaporates, depends on the geometry, thermal conditions and thermodynamical properties of the liquid and the surrounding gas. The numerical prediction of the evaporation rate in complex geometries is a part of recent research \cite{kubochkin_capillary-driven_2022}.
    
\begin{figure}[bt]
        \centering
        \begin{subfigure}{0.48\linewidth}
            \centering
            \fontsize{11pt}{13pt}\selectfont
            \def\svgwidth{0.5\linewidth}
            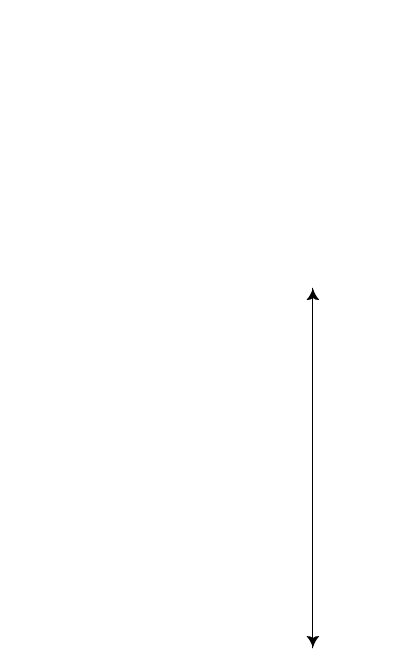
            \caption{Round capillary.}
            \label{fig:RoundCapillarySketch} 
        \end{subfigure}
        %\hspace{0.5cm}
        \begin{subfigure}{0.48\linewidth}
            \centering
            \fontsize{11pt}{13pt}\selectfont
            \def\svgwidth{0.5\linewidth}
            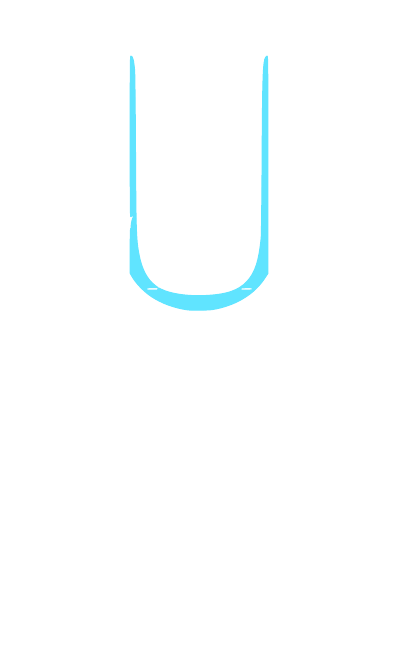
            \caption{Square capillary.}
            \label{fig:SquareCapillarySketch}
        \end{subfigure}
        \caption[Sketch of round and square capillary sections.]{Section through the center of regular shaped capillaries with a contact angle of $\theta<45$° and a mean diffusion length of $\bar{\chi}>0$. For round capillaries (a), the meniscus shape is an approximate spherical cap. For a contact angle of $\theta<45$°, liquid fingers develop in the corners of square capillaries (b).}
        \label{fig:CapillarySketch}   
    \end{figure}

    First investigations on the evaporation of water from capillaries go back to Stefan in 1871 \cite{stefan_uber_1871}. He proposed an analytical model to calculate the evaporation rate from round capillaries based on Fick's law of diffusion. Since then, experiments were conducted which confirm Stefan's model \cite{camassel_evaporation_2005,schweigler_evaporation_2018}. The evaporation rate in Stefan's model depends on the mean diffusion length $\bar{\chi}$ which describes the distance between the open capillary end and the mean meniscus position as depicted in \cref{fig:CapillarySketch}. Although this model is in good agreement with experimental results, it does not take the meniscus shape into account and assumes a flat surface.

    In 1805, Young formulated a model to describe the change in pressure $\Delta p$ at a fluid-fluid interface, known as the capillary pressure, which depends on the interface curvature $\kappa$ \cite{royal_society_great_britain_philosophical_1805}. In the same year, this model was mathematically described by Laplace yielding the Young-Laplace equation \cite{de_laplace_traite_1805}. Concus and Finn studied the shape of liquid surfaces in a wedge and showed that the Young-Laplace equation only has a bounded solution if $\theta>\pi/2-\alpha$, where $\theta$ is the contact angle as depicted in \cref{fig:CapillarySketch} and $\alpha$ is the half wedge angle \cite{concus_behavior_1969}.

    Violating the condition by Concus and Finn leads to the formation of liquid fingers in the corners of the capillary as shown in \cref{fig:SquareCapillarySketch}. Depending on corner geometry and liquid properties, the liquid fingers may reach up until the end of the capillary \cite{ransohoff_laminar_1988, chauvet_depinning_2010} which results in a local reduction of diffusion length in the vicinity of the liquid fingers \cite{schweigler_evaporation_2018}. The reduced diffusion length in turn results in an enhanced evaporation, which could be confirmed in various experiments \cite{keita_drying_2016, schweigler_evaporation_2018}.

    Ransohoff and Radke performed numerical studies on non-circular capillaries to calculate corner liquid flow. They introduced a dimensionless flow resistance to account for corner angle, corner roundedness, surface shear viscosity and contact angle \cite{ransohoff_laminar_1988}. Chauvet et al. built up on their findings and showed that corner roundedness has a significant influence on the liquid finger geometry and might lead to depinning of the fingers from the top of the capillary \cite{chauvet_depinning_2010}. Additionally, Chauvet et al. separated the drying process of capillaries into the three phases constant rate period (CRP), falling rate period (FRP) and receeding front period (RFP) depending on the free diffusion length \cite{chauvet_three_2009}. CRP is characterized by the liquid-gas interface touching the open end of the capillary providing a nearly constant evaporation rate. Further evaporation leads to depinning of the liquid fingers and a drop in evaporation rate. In experiments, the depinning tends to coincide with the initiation of the FRP. Nevertheless, Keita et al. \cite{keita_drying_2016} showed for a rectangular channel where gravity was neglected, that liquid fingers might also be present in FRP. During RFP the meniscus shape can be assumed to stay constant, leading to an internal evaporation front receding inside the capillary \cite{chauvet_three_2009}.
    
    Two centuries after Stefan's findings, Camassel et al. proposed a simplified analytical model for square capillaries that neglects gravity and viscous effects to calculate the evaporation rate \cite{camassel_evaporation_2005}. A few years later, Yiotis et al. developed an analytical model that also takes gravity and viscous effects into account \cite{yiotis_analytical_2012} to determine the three-phase drying process in non-circular capillaries which was previously proposed by Chauvet et al. \cite{chauvet_three_2009, chauvet_depinning_2010}. Since analytical solutions are mostly limited to simple geometries, such as round capillaries, numerical models were developed that enable prediction of evaporation in arbitrary shaped geometries.
    
    A major problem in common numerical methods that enable a full resolution of the free liquid surface, e.g. Volume-of-Fluid (VoF) or Front-Tracking, is the numerical handling of the mass source term at the liquid-gas interface due to evaporation in combination with the interface movement \cite{scapin_volume--fluid_2020, kubochkin_capillary-driven_2022, hardt_evaporation_2008}. Irfan and Muradoglu developed a model to incorporate concentration-driven evaporation for the front-tracking method \cite{irfan_front_2017}. Their model is tested for 2D cases but they state that it could be easily adapted to 3D cases as well. Nonetheless, the Front-Tracking method is difficult to parallelize, thus it's hardly applicable to complex simulations with several Million grid cells \cite{farooqi_communication_2019}. For the VoF approach a volume fraction field is used which allows interface capturing. The evaporation mass source term results in a non-divergence-free volume fraction field which tends to produce artificial spurious currents. To counteract this unintended behavior due to jump conditions at the interface, several methods were developed to dampen spurious currents \cite{hardt_evaporation_2008, schlottke_direct_2008, palmore_volume_2019, scapin_volume--fluid_2020}. In all previously mentioned works, the Navier-Stokes equations are solved in combination with an energy equation which results in a high computational effort. However, Schweigler et al. used an Allen-Cahn phase-field approach where only an energy functional is solved to predict evaporation behavior in capillaries \cite{schweigler_evaporation_2018}. Although this method yields good results for arbitrary capillary shapes, it strongly relies on numerical parameters, gathered from calibration with experimental data.

    Experimental investigations from Schweigler et al. \cite{schweigler_analyse_2017} show that the evaporation of $V_0\approx 40\unit{\upmu l}$ from a round capillary with 1~mm diameter can take more than 5 days. Thereby, significant time scale differences between evaporation and diffusion occur \cite{sultan_evaporation_2004} that must be resolved by conventional methods with time steps in the order of $\Delta t\approx10^{-6}\unit{s}$. Considering small time steps and large time periods to be simulated, it is neither computationally nor temporally feasible to calculate the entire drying process of a capillary. Consequently, we present a decoupled approach where the fluid-fluid interface simulation is separated from the diffusion simulation. The goal of this work is to develop a generalized approach to calculate the evaporation rate from liquid to gas for various geometries, given specific thermodynamical boundary conditions.  To the best of our knowledge there is no similar method in literature with a comparable speed while still providing accurate results.
    %\input{theoreticalBackground}
    
    % Methodology
    \section{Methodology}
\label{sec:methodology}
    For the proposed quasi-steady approach, gas flow above and inside cavities is neglected to simplify the underlying Convection-Diffusion equation. Additionally, isothermal conditions are assumed such that the evaporation rate of liquid in a gas environment can be calculated for time steps $t_i$ individually according to
    ~
    \begin{equation}
        \label{eq:evapRateGeneral}
        \dot{m}\ped{ev}(t_i)=\dot{m}\ped{ev}\left(\rho\ped{g},D,Y\ped{v,sat},\frac{\diff Y}{\diff \bm{n}},A_{\Sigma}(\Omega,\bar{\chi})\right)
    \end{equation}
    
    where $\rho\ped{g}$ is the gas density, $D$ is the diffusion coefficient of the evaporating species in the inert species, $Y\ped{v,sat}$ is the saturation vapor mass fraction, $\diff Y/\text{d}\bm{n}$ is the partial derivative of vapor mass fraction with respect to the interface normal vector $\bm{n}$ and $A_{\Sigma}$ is the liquid-gas interface surface area which depends on the capillary geometry $\Omega$ and the mean diffusion length $\bar{\chi}$. \color{\revonecolor}The latter describes the liquid filling height of a cavity in case of a flat surface (see \cref{fig:CapillarySketch}) and can also be considered as the distance, vapor molecules have to be transported out of the capillary. It is calculated as

    \begin{equation}
        \label{eq:meanDiffusionLengthDef}
        \bar{\chi} = h\ped{c}-h\ped{l} = \frac{m\ped{ev}}{A\ped{c}\rho\ped{l}}
    \end{equation}

    with the capillary height $h\ped{c}$ and the actual liquid filling height $h\ped{l}$ according to \cref{fig:CapillarySketch}. Further, $m\ped{ev}$ is the evaporated liquid mass, $A\ped{c}$ is the capillary cross-section and $\rho\ped{l}$ is the liquid density. In addition to the spatial meaning, the mean diffusion length also has a temporal meaning, since $\bar{\chi}$ increases with time as the water evaporates and the meniscus recedes inside the capillary.\color{black} %In order to simplify the simulation setup and enhance comparability to experimental results, $\bar{\chi}$ is used throughout this work. 
    
    The times $t_i$, at which the evaporation rate is evaluated, are unknown prior to the simulation. Thus, the function $f\,:\,\bar{\chi}\xrightarrow{}\dot{m}\ped{ev}$ is solved which can be transformed into $g\,:\, m\ped{ev}\xrightarrow{}\dot{m}\ped{ev}$ by substituting $\bar{\chi}$ using \cref{eq:meanDiffusionLengthDef}. Since $\dot{m}\ped{ev}=\diff m\ped{ev}/\text{d}t$, the time evaluates to
    ~
    \begin{equation}
        \label{eq:timeDependence}
        t_i = \int_M \frac{1}{\dot{m}\ped{ev}}\diff m\ped{v}.
    \end{equation}

    Utilizing \cref{eq:timeDependence}, the water mass $m\ped{w}(t_i)$ is determined as 
    ~
    \begin{equation}
        m\ped{w}(t_i)=m\ped{w}(t_i=0)-m\ped{ev}(t_i).
    \end{equation}

    In order to determine the function $f\,:\,\bar{\chi}\xrightarrow{}\dot{m}\ped{ev}$, the following three-step method is proposed:

    \begin{enumerate}
        \item \textbf{Surface formation simulation}: The liquid surface shape $A_{\Sigma}(\Omega, \bar{\chi})$ is determined with Surface Evolver (SE). It is initialized as a plane and treated as a discrete differential geometry with movable vertices. The total surface energy is then minimized using a gradient-descent approach.
        \item \textbf{Pre-processing}: The capillary geometry is extended by an environment domain above the open end of the capillary. The liquid surface from SE divides the resulting domain into a liquid and a gas domain. By utilizing the OpenFOAM (OF) meshing utility snappyHexMesh (sHM), only the gas domain is meshed and the liquid domain is neglected.
        \item \textbf{Diffusion simulation}: The evaporation rate is determined using OF which solves the Laplace equation $\nabla\cdot(\rho\ped{g}D\nabla Y_v)$.
    \end{enumerate}

\begin{figure}[tb]
        \centering
        \fontsize{10pt}{13pt}\selectfont
        \def\svgwidth{\textwidth}
        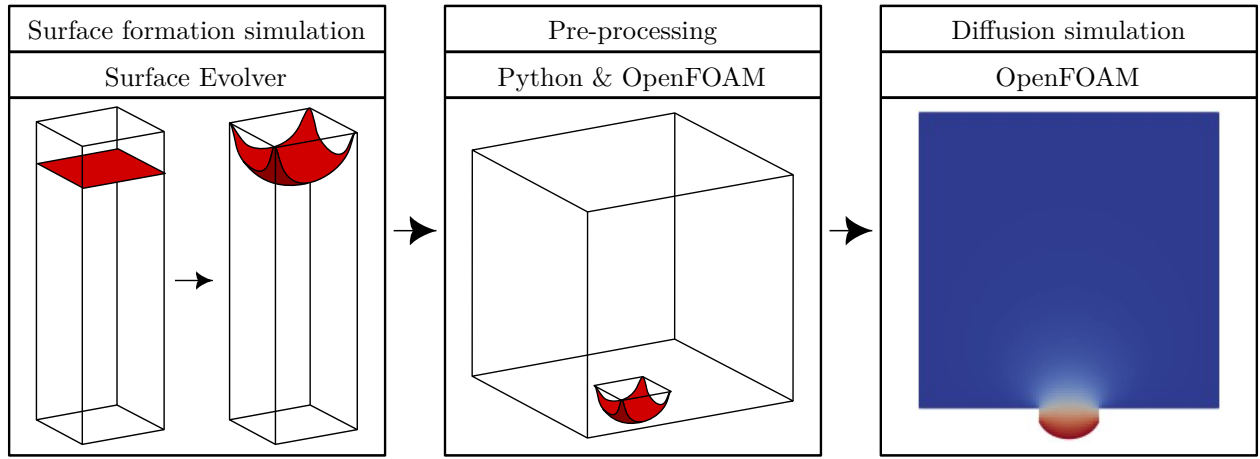
        \caption[Flowchart of the proposed three-step method.]{Schematic process of the proposed three-step method to determine evaporation rate in cavities. (left) Flat plane liquid initialization and surface formation simulation. (center) Extension of cavity by an environmental domain, slicing at the simulated liquid surface and extraction of gas domain. (right) Vapor-in-gas diffusion simulation and evaporation rate calculation by integrating the vapor mass fraction flux over the environment boundary.}
        \label{fig:flowchart} 
    \end{figure}

    The previously described three-step process is depicted in \cref{fig:flowchart}. The following sections describe the separate steps in more detail. The complete source code of the presented method is uploaded to the Bosch Research GitHub \footnote{\url{https://github.com/boschresearch/sepMultiphaseFoam/tree/publications/novelSimulationApproachForEvaporationProcesses}}.

    \subsection{Surface formation simulation}
        \label{sec:interfaceSim}
        The surface shape $A_{\Sigma}$ is treated as a function of the crevice geometry $\Omega$ and the mean diffusion length $\bar{\chi}$ as given in \cref{eq:evapRateGeneral}. For the surface formation simulation, the SE is used which was developed by Brakke in 1992 \cite{brakke_surface_1992}. SE allows the user to create own geometries by defining vertices, edges and facets and then calculate the interface shape which is subject to different boundary conditions such as contact angle or surface tension. The algorithm behind SE is a gradient-descent method which minimizes the total surface energy of a given surface \cite{brakke_surface_1992}. SE further provides features to manipulate the surface mesh such as refinement, equiangulation or vertex averaging which are described in the SE manual in more detail \cite{brakke_surface_2013}.

        Since SE doesn't provide a convergence algorithm, the solution depends on the sequence of iteration and surface manipulation steps. To prevent this, the convergence algorithm from the \textit{Surface Evolver - Fluid Interface Tool} (SE-FIT\textsuperscript{\textregistered}) is used which is developed at the Portland State University \cite{chen_introducing_2011}. In order to serve the presented purpose, the algorithm settings are tuned which is further discussed in \cref{appSec:artificialDepinning}.

        \subsubsection*{Simulation setup}

        In this work, capillaries with various cross section shapes are considered. \revone{For all cases, a parameterized input file for SE is created with the mean diffusion length $\bar{\chi}$ as an input parameter.} All geometries are centered around the origin and the bottom of the capillary is aligned with the plane at $z=0$ as visualized in \cref{fig:CapillarySketch}.

\begin{figure}[tb]
            \centering
            \fontsize{11pt}{13pt}\selectfont
            \def\svgwidth{0.27\textwidth}
            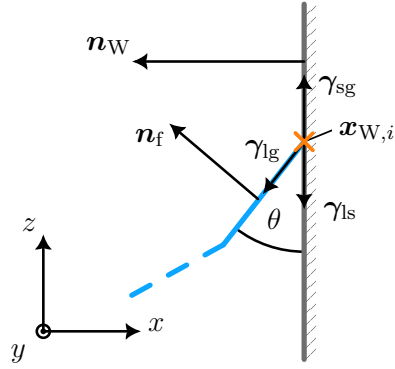
            \caption[Sketch of fluid surface at the wall.]{2D sketch of an arbitrary fluid surface facet at the wall with the corresponding surface energies $\gamma\ped{lg},\,\gamma\ped{sg},\,\gamma\ped{ls}$ between solid, gas and liquid. $\bm{n}\ped{W}$ and $\bm{n}\ped{f}$ are the wall and facet normal vectors and $\bm{x}_{\mathrm{W,}i}$ are the coordinates of a wall vertex.}
            \label{fig:SE_SurfaceEnergies} 
        \end{figure}

        To simplify the file setup and reduce simulation runtime, completely wetted wall areas are omitted in the setup, thus leaving only the free fluid surface as depicted in \cref{fig:SE_SurfaceEnergies}.
        
        In the surface formation simulations, the liquid surface is initialized as a plane at $z=h\ped{l}$ where $h\ped{l}$ is the liquid filling height \revone{(see \cref{fig:CapillarySketch} and \cref{appSec:SEIterationProcess})} and then discretized as facets, edges and vertices. In order to comply with the contact angle boundary condition, the surface must be deformed. The surface deformation is done by calculating a vertex displacement vector depending on the local surface energy contribution, thus energy constraints must be implemented at the walls. The total liquid-gas surface energy at the walls is
        ~
        \begin{equation}
            \label{eq:wallSurfaceEnergyAreaIntegral}
            E\ped{\gamma,W} = \iint_{A\ped{W}} \bm{\gamma}\ped{lg}\cdot\text{d}\bm{A}
        \end{equation}

        with the wall area $A\ped{W}$ and the surface energy 
        ~
        \begin{equation}
            \bm{\gamma}\ped{lg}=\sigma\cos\theta\begin{pmatrix}x_{\mathrm{W,}i}\\y_{\mathrm{W,}i}\\0\end{pmatrix}
        \end{equation}
        
        between liquid and gas where $\sigma$ is the surface tension coefficient. By applying Stoke's theorem, the integral in \cref{eq:wallSurfaceEnergyAreaIntegral} can be transformed into a line integral around the three phase contact line $\bm{s}\ped{CL}$ of the surface. This calls for a divergenceless vector field $\bm{\gamma}\ped{lg}$, thus
        ~
        \begin{equation}
            \label{eq:divergencelessRequirement}
            \bm{\gamma}\ped{lg} = \nabla\times \bm{F}
        \end{equation}

        where $\bm{F}$ is a vector potential \cite{brakke_surface_2013}. Substituting $\gamma\ped{lg}$ from \cref{eq:divergencelessRequirement} in \cref{eq:wallSurfaceEnergyAreaIntegral} yields
        ~
        \begin{equation}
            E\ped{\gamma,W} = \oint_{\bm{s}\ped{CL}} \bm{F}\cdot \text{d}\bm{s}.
        \end{equation}

        In order to satisfy \cref{eq:divergencelessRequirement}, $\bm{F}$ must be specified. In this work, the vector potential is set to 
        ~
        \begin{equation}
            \bm{F} = \sigma\cos\theta\begin{pmatrix}z_{\mathrm{W,}i}\\-z_{\mathrm{W,}i}\\0\end{pmatrix}.
        \end{equation}

        The initial filling height $h\ped{l}$ depends on the mean diffusion length as $h\ped{l}=h\ped{c}-\bar{\chi}$ (see \cref{eq:meanDiffusionLengthDef}), thus the volume at the start of simulation is given as $V_0=h\ped{l}A\ped{c}$. Since the liquid is assumed to be incompressible, its volume must be conserved. In order to comply with volume conservation, the volume error at each iteration is calculated as
        ~
        \begin{equation}
            \label{eq:volumeConservation}
            \varepsilon\ped{V} = V_0-\iint_{\Sigma}\begin{pmatrix}0\\0\\z_i\end{pmatrix}\cdot\diff\bm{\sigma}
        \end{equation}

        where $\bm{\sigma}$ is a facet of the surface $\Sigma$ and $z_i$ are the vertices z coordinates. The volume error then contributes to the surface energy loss function.

        Further contributions to the surface energy calculation through gravity and surface tension in the bulk are taken care of by the SE and are described in the SE manual in detail \cite{brakke_surface_2013}.      

    \subsection{Pre-processing}
\begin{figure}[tb]
            \centering
            \fontsize{10pt}{10pt}\selectfont
            \def\svgwidth{\textwidth}
            %% Creator: Inkscape 1.3.2 (091e20e, 2023-11-25, custom), www.inkscape.org
%% Accompanies image file 'Figure_4.pdf' (pdf, eps, ps)
\begingroup%
  \makeatletter%
  \providecommand\color[2][]{%
    \errmessage{(Inkscape) Color is used for the text in Inkscape, but the package 'color.sty' is not loaded}%
    \renewcommand\color[2][]{}%
  }%
  \providecommand\transparent[1]{%
    \errmessage{(Inkscape) Transparency is used (non-zero) for the text in Inkscape, but the package 'transparent.sty' is not loaded}%
    \renewcommand\transparent[1]{}%
  }%
  \providecommand\rotatebox[2]{#2}%
  \newcommand*\fsize{\dimexpr\f@size pt\relax}%
  \newcommand*\lineheight[1]{\fontsize{\fsize}{#1\fsize}\selectfont}%
  \ifx\svgwidth\undefined%
    \setlength{\unitlength}{739.8837833bp}%
    \ifx\svgscale\undefined%
      \relax%
    \else%
      \setlength{\unitlength}{\unitlength * \real{\svgscale}}%
    \fi%
  \else%
    \setlength{\unitlength}{\svgwidth}%
  \fi%
  \global\let\svgwidth\undefined%
  \global\let\svgscale\undefined%
  \makeatother%
  \begin{picture}(1,0.32083936)%
    \lineheight{1}%
    \setlength\tabcolsep{0pt}%
    \put(0,0){\includegraphics[width=\unitlength,page=1]{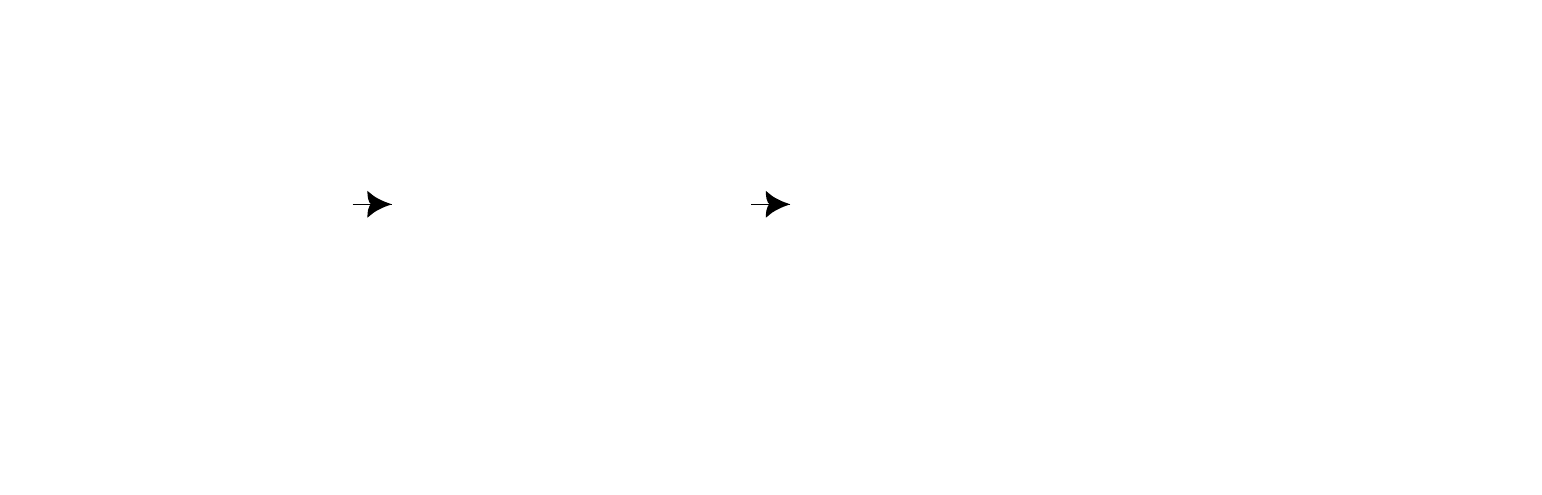}}%
    \put(0.11137712,0.28724216){\color[rgb]{0,0,0}\makebox(0,0)[t]{\lineheight{1.25}\smash{\begin{tabular}[t]{c}Model capillary\end{tabular}}}}%
    \put(0,0){\includegraphics[width=\unitlength,page=2]{Figure_4.pdf}}%
    \put(0.36958945,0.29670668){\color[rgb]{0,0,0}\makebox(0,0)[t]{\lineheight{1.25}\smash{\begin{tabular}[t]{c}Environment\\extension\end{tabular}}}}%
    \put(0,0){\includegraphics[width=\unitlength,page=3]{Figure_4.pdf}}%
    \put(0.62885875,0.29817561){\color[rgb]{0,0,0}\makebox(0,0)[t]{\lineheight{1.25}\smash{\begin{tabular}[t]{c}Load surface\\\& scaling\end{tabular}}}}%
    \put(0,0){\includegraphics[width=\unitlength,page=4]{Figure_4.pdf}}%
    \put(0.88752635,0.29834693){\color[rgb]{0,0,0}\makebox(0,0)[t]{\lineheight{1.25}\smash{\begin{tabular}[t]{c}Domain extraction\\\& Meshing\end{tabular}}}}%
    \put(0,0){\includegraphics[width=\unitlength,page=5]{Figure_4.pdf}}%
  \end{picture}%
\endgroup%

            \caption[Flowchart of the pre-processing step.]{Schematic pre-processing step to split the capillary at the simulated 2D surface and mesh the resulting gas domain.}
            \label{fig:PreProcessing} 
        \end{figure}

        Since SE calculates a 2D surface in 3D space, the surface must be embedded into a volume mesh in order to simulate diffusion in the following three-dimensional vapor simulation step. Therefore, the capillary geometry is built in OF and extended by an environment over the open end of the capillary as depicted in \cref{fig:PreProcessing}.

        In order to create the curved meniscus shape as a lower boundary, the empty capillary is modeled first. As a second step the evolved interface from SE is scaled up by 1\% in $x$ and $y$ direction to ensure that the interface completely intersects the outer capillary walls. The interface then splits the capillary into a liquid and a gas domain. As a third step, sHM is used to create a volume mesh for the upper gas domain.

    \subsection{Vapor diffusion simulation}
        \label{sec:diffusionSim}
\begin{figure}[tb]
            \centering
            \fontsize{11pt}{13pt}\selectfont
            \def\svgwidth{0.6\linewidth}
            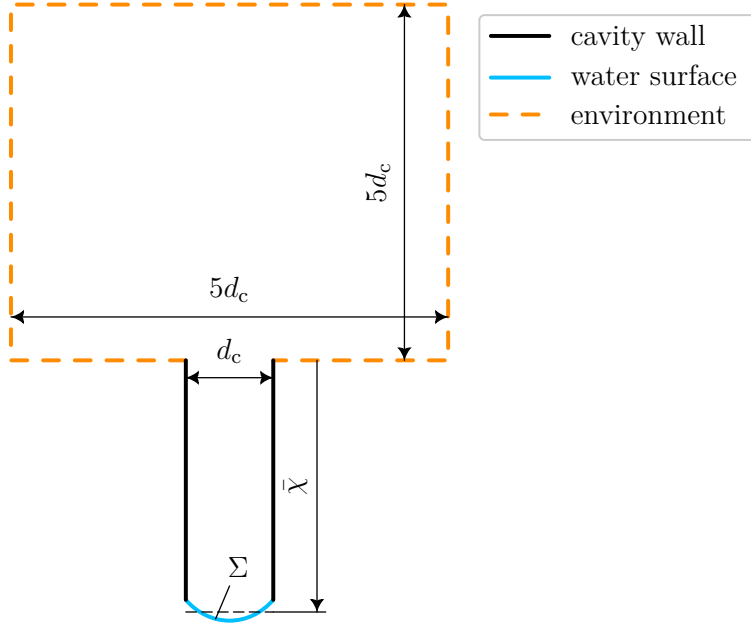
            \caption[Domain considered for vapor diffusion simulation.]{Air domain of a round capillary with an extended environment domain and the simulated water surface from SE.}
            \label{fig:CapillarySketch_airDomain}   
        \end{figure}

        For the vapor diffusion simulation, the liquid phase is completely neglected leaving only the gas domain as depicted in \cref{fig:CapillarySketch_airDomain}. The transport of multiple miscible species in a domain is governed by the convection-diffusion equation
        ~
        \begin{equation}
            \label{eq:ConvDiffEquation}
            \frac{\partial\rho_i}{\partial t}+\nabla\cdot(\rho_i\bm{v})=-\nabla\cdot\left(\rho_i D \nabla Y_i\right)
        \end{equation}

        for each species $i$. In \cref{eq:ConvDiffEquation}, $\rho_i$ is the species density and $\bm{v}=(v_x,v_y,v_z)^{\text{T}}$ is the velocity. In the following, a two-species system is assumed, consisting of an inert specie (index in) and a vapor specie (index v), i.e. no chemical reactions take place. Under this assumption the inert specie is omitted, since its mass fraction can be calculated as $Y\ped{in}=1-Y\ped{v}$.

        The inert gas at the interface is assumed to be constantly saturated. The vapor saturation mass fraction depends on the vapor saturation pressure which can be calculated using the Antoine equation
        ~
        \begin{equation}
            \label{eq:saturationPressure}
            p\ped{v,sat} = \exp\left(\kappa\ped{A}-\frac{\kappa\ped{B}}{T+\kappa\ped{C}}\right) 
        \end{equation}

        with the Antoine coefficients $\kappa\ped{A},\kappa\ped{B},\kappa\ped{C}$. The mole fraction of vapor in the gas mixture is given by
        ~
        \begin{equation}
            \label{eq:moleFraction}
            X\ped{v} = \frac{p\ped{v}}{p\ped{tot}}
        \end{equation}

        where $p\ped{v}$ is the vapor partial pressure. The mole fraction can then be converted to a mass fraction using
        ~
        \begin{equation}
            \label{eq:massFraction}
            Y\ped{v} = \frac{X\ped{v}M\ped{v}}{X\ped{v}M\ped{v}+(1-X\ped{v})M\ped{in}}
        \end{equation}

        with the molar masses $M\ped{v},M\ped{in}$ of vapor and inert specie respectively. The domain is assumed to be isothermal, hence evaporative cooling is not considered and no energy equation needs to be solved. In this case, the saturation vapor mass fraction $Y\ped{v,sat}$ is constant at the interface boundary. 
        
        To determine the quasi-steady solution of \cref{eq:ConvDiffEquation}, the temporal dependence ${\diff/\diff t=0}$. Air flow over the capillary is also neglected such that $\bm{v}=0$. Thus, the vapor in the domain is only subject to molecular diffusion. For liquid and gas at rest, the evaporation time scale $\bar{\chi}\rho\ped{l}/\dot{m}\ped{ev}$ is much smaller than the diffusion time scale $\bar{\chi}^2/D$. According to Sultan et al. \cite{sultan_evaporation_2004}, this further allows the assumption of a quasi-steady interface and that the distribution of vapor in air $Y\ped{v}$ is a solution of the Laplace equation
        ~
        \begin{equation}
            \label{eq:laplace}
            -\nabla\cdot\left(\rho_i D \nabla Y_i\right)=0.
        \end{equation}

        \subsubsection*{Simulation setup}

        In the gas domain, only the vapor specie is considered and the inert specie is omitted as stated before. For all boundaries in \cref{fig:CapillarySketch_airDomain} and simulations, boundary conditions must be specified. They are listed in \cref{tab:boundaryConditions}.

        \begin{table}[tb]
            \centering
            \begin{tabular}{lll}
                \hline
                boundary name       & boundary type     & value \\
                \hline
                liquid surface      & Dirichlet         & $Y\ped{v,sat}$\\
                capillary walls     & Neumann           & $\nabla Y\ped{v}=0$\\
                environment         & Dirichlet         & $Y\ped{v,\infty}$\\
                \hline
            \end{tabular}
            \caption{Vapor mass fraction boundary conditions for diffusion simulations.}
            \label{tab:boundaryConditions}
        \end{table}

        To solve the Laplace \cref{eq:laplace} for the given domain, a customized version of the laplacianFoam solver from OF is used, which is called steadyMolecularDiffusionFoam. The solver is adapted to calculate the steady-state Laplace solution by dropping all time-dependent terms in the \textit{laplacianFoam} code. The numerical parameters that are used across all simulations are listed in \cref{tab:numericalParams}.

        \begin{table}[tb]
            \centering
            \begin{tabular}{ll}
                \hline
                parameter/setting   & value \\
                \hline
                solver              & PCG\\
                preconditioner      & DIC\\
                tolerance           & $10^{-6}$\\
                relTol              & 0\\
                laplacianSchemes    & Gauss linear corrected\\
                $N\ped{iter}$       & 10\\
                \hline
            \end{tabular}
            \caption{\revtwo{Numerical settings and parameters that are used for the diffusion simulations in OpenFOAM. These parameters are generally applicable to all cases discussed in this study.}}
            \label{tab:numericalParams}
        \end{table}
    
    % Results and discussion
    \section{Results and discussion}
    \label{sec:resultsDiscussion}
    
    Simulations are conducted for the evaporation of water in air for round and square capillaries according to the experiments from Schweigler \cite{schweigler_evaporation_2018}. A height of $h\ped{c}=50$~mm is used for all capillaries as in the experiments. The temperature $T=23$~°C and the pressure $p=101325$~Pa are constant for all simulations. The environmental mass fraction $Y\ped{v,\infty}$ is set according to a relative humidity of $\varphi=p\ped{v}/p\ped{v,sat}=17$ \%. The temperature dependence of density $\rho\ped{g}$ and diffusion coefficient $D$ are incorporated using the ideal gas law for the humid air gas mixture and the Fuller method \cite{fuller_diffusion_1969, vdi_e_v_vdi-warmeatlas_2013} respectively. The thermodynamical properties of water and air under the given conditions are listed in \cref{tab:FluidProperties}.

    \begin{table}[tb]
        \centering
        \begin{tabular}{lcccc}
            \hline
            Property                                            & Water             & Air                   & Unit          & Reference\\
            \hline
            Environment vapor mass fraction $Y\ped{v,\infty}$   & -                 & $2.984$               & \unit{g/kg}   & \cref{eq:saturationPressure,eq:moleFraction,eq:massFraction}\\
            Saturated vapor mass fraction $Y\ped{v,sat}$        & -                 & $17.71$               & \unit{g/kg}   & \cref{eq:saturationPressure,eq:moleFraction,eq:massFraction}\\
            Density $\rho$                                      & $997.5$           & $1.184$               & \unit{kg/m^3} & \cite{vdi_e_v_vdi-warmeatlas_2013}\\
            Surface tension $\sigma$                            & \multicolumn{2}{c}{$0.072$}               & \unit{N/m}    & \cite{vdi_e_v_vdi-warmeatlas_2013}\\
            Diffusion coefficient $D$                           & \multicolumn{2}{c}{$24.75$}               & \unit{mm^2/s}  & \cite{fuller_diffusion_1969, vdi_e_v_vdi-warmeatlas_2013}\\
            Contact angle $\theta$ (with borosilicate glass)    & \multicolumn{2}{c}{$20$}                  & $^{\circ}$    & \cite{schweigler_analyse_2017}\\
            Antoine coefficient $\kappa\ped{A}$                 & \multicolumn{2}{c}{23.4751}               & -             & \cite{reid_properties_1987}\\
            Antoine coefficient $\kappa\ped{B}$                 & \multicolumn{2}{c}{3978.205}              & \unit{K}      & \cite{reid_properties_1987}\\
            Antoine coefficient $\kappa\ped{C}$                 & \multicolumn{2}{c}{-39.801}               & \unit{K}      & \cite{reid_properties_1987}\\
            \hline
        \end{tabular}
        \caption{Fluid properties of water and air for $T=23$\unit{^\circ C} and $p=101325$\unit{Pa}. Intermediate values are linearly interpolated.}
        \label{tab:FluidProperties}
    \end{table}
    
    The evaporation rate is calculated and used as a measure to compare the simulative results to the experimental results. It is evaluated as the area integrated flux of vapor over the environment boundaries $\Omega$ such that
    ~
    \begin{equation}
        \label{eq:integratedMassFraction}
        \dot{m}\ped{ev} = \rho\ped{g}D\int_{\Omega}\nabla Y\ped{v} \diff\omega.
    \end{equation}
    
    \subsection{Round capillaries}
        \label{sec:ResultsRoundCapillaries}
        First, two round capillaries with an inner diameter $d\ped{c}$ of $1\unit{mm}\text{ and }10\unit{mm}$ (cases R1, R10) are investigated. Simulations are conducted for a mean diffusion length $\bar{\chi}$ in the range of $\bar{\chi}=0.2-20\unit{mm}$. In order to better resolve high evaporation rate gradients for $\bar{\chi}\xrightarrow{}0$, the stepwidths $\Delta\bar{\chi}$ are unevenly distributed with 
        ~
        \begin{equation}
            \Delta\bar{\chi}=\begin{cases}
                0.2\unit{mm} &\text{for}\: 0\unit{mm}<\bar{\chi}\leq 2\unit{mm},\\
                1\unit{mm}   &\text{for}\: 2\unit{mm}<\bar{\chi}\leq 5\unit{mm},\\
                2.5\unit{mm} &\text{for}\: 5\unit{mm}<\bar{\chi}\leq 20\unit{mm}.
            \end{cases}
            \label{eq:meanDiffusionLengths}
        \end{equation}
        
\begin{figure}[tb]
            \centering
            \includegraphics[width=0.07\linewidth]{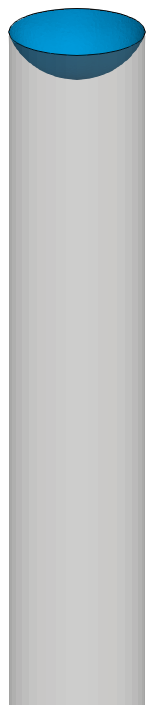}
            \includegraphics[width=0.07\linewidth]{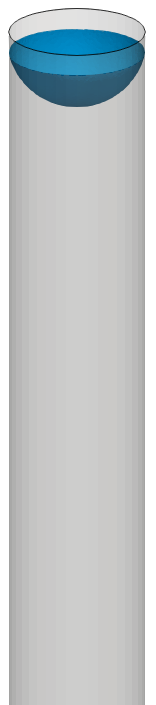}
            \includegraphics[width=0.07\linewidth]{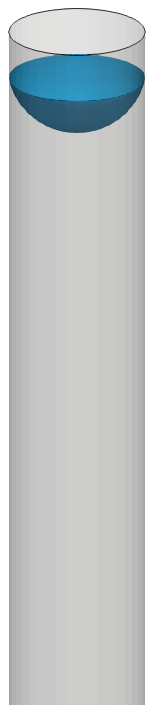}
            \includegraphics[width=0.07\linewidth]{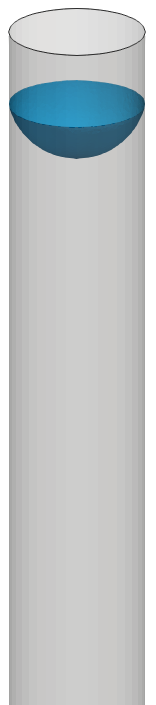}
            \includegraphics[width=0.07\linewidth]{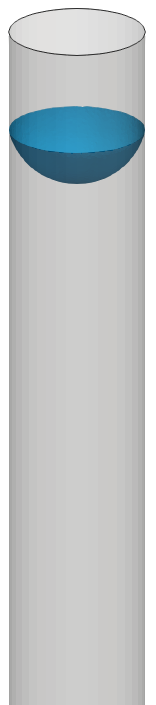}
            \includegraphics[width=0.07\linewidth]{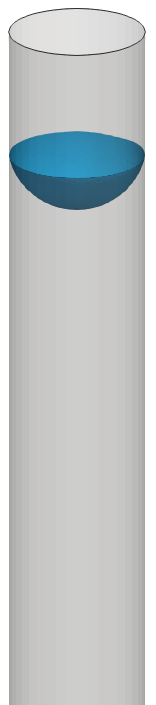}
            \includegraphics[width=0.07\linewidth]{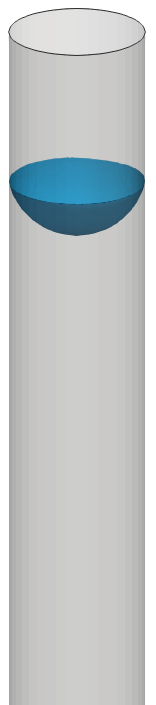}
            \includegraphics[width=0.07\linewidth]{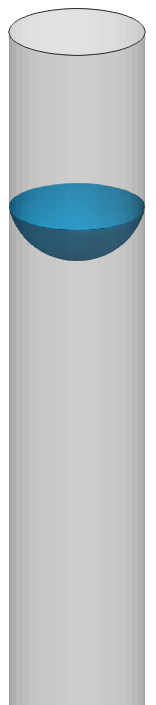}
            \includegraphics[width=0.07\linewidth]{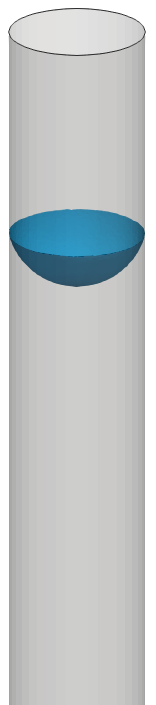}
            \includegraphics[width=0.07\linewidth]{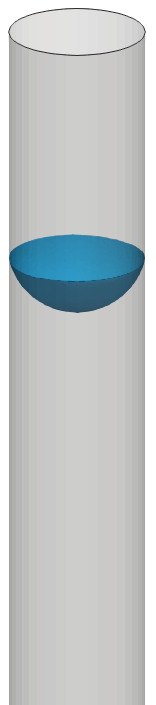}
            \includegraphics[width=0.07\linewidth]{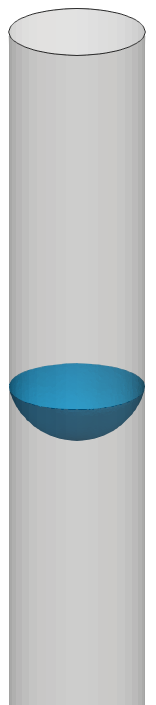}
            \includegraphics[width=0.07\linewidth]{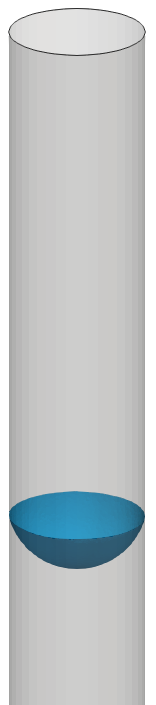}
            \includegraphics[width=0.07\linewidth]{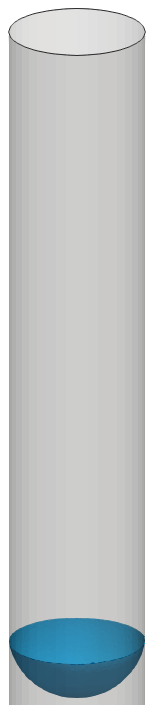}
            \par\vspace{2mm}
            \fontsize{12pt}{12pt}\selectfont
            \def\svgwidth{\linewidth}
            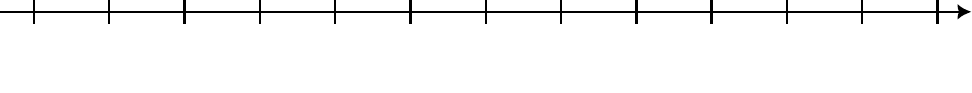
            \caption[Fluid surface of round capillary as simulated in Surface Evolver.]{Simulated fluid surface from SE for the round capillary with a diameter of \revone{$d\ped{c}=1\unit{mm}$}, height of $h\ped{c}=50\unit{mm}$ and a contact angle of $\theta=20^{\circ}$.}
            \label{fig:SEresultsR1}
        \end{figure}

        For each diffusion length, SE simulates the equilibrium fluid surface. The results from SE for the R1 case are shown in \cref{fig:SEresultsR1}.
        
\begin{figure}[tb]
            \color{\revonecolor}
            \centering
            \fontsize{10pt}{10pt}\selectfont
            \def\svgwidth{\linewidth}
            %% Creator: Inkscape 1.3.2 (091e20e, 2023-11-25, custom), www.inkscape.org
%% Accompanies image file 'Figure_7.pdf' (pdf, eps, ps)
\begingroup%
  \makeatletter%
  \providecommand\color[2][]{%
    \errmessage{(Inkscape) Color is used for the text in Inkscape, but the package 'color.sty' is not loaded}%
    \renewcommand\color[2][]{}%
  }%
  \providecommand\transparent[1]{%
    \errmessage{(Inkscape) Transparency is used (non-zero) for the text in Inkscape, but the package 'transparent.sty' is not loaded}%
    \renewcommand\transparent[1]{}%
  }%
  \providecommand\rotatebox[2]{#2}%
  \newcommand*\fsize{\dimexpr\f@size pt\relax}%
  \newcommand*\lineheight[1]{\fontsize{\fsize}{#1\fsize}\selectfont}%
  \ifx\svgwidth\undefined%
    \setlength{\unitlength}{1016.75949457bp}%
    \ifx\svgscale\undefined%
      \relax%
    \else%
      \setlength{\unitlength}{\unitlength * \real{\svgscale}}%
    \fi%
  \else%
    \setlength{\unitlength}{\svgwidth}%
  \fi%
  \global\let\svgwidth\undefined%
  \global\let\svgscale\undefined%
  \makeatother%
  \begin{picture}(1,0.49439497)%
    \lineheight{1}%
    \setlength\tabcolsep{0pt}%
    \put(0,0){\includegraphics[width=\unitlength,page=1]{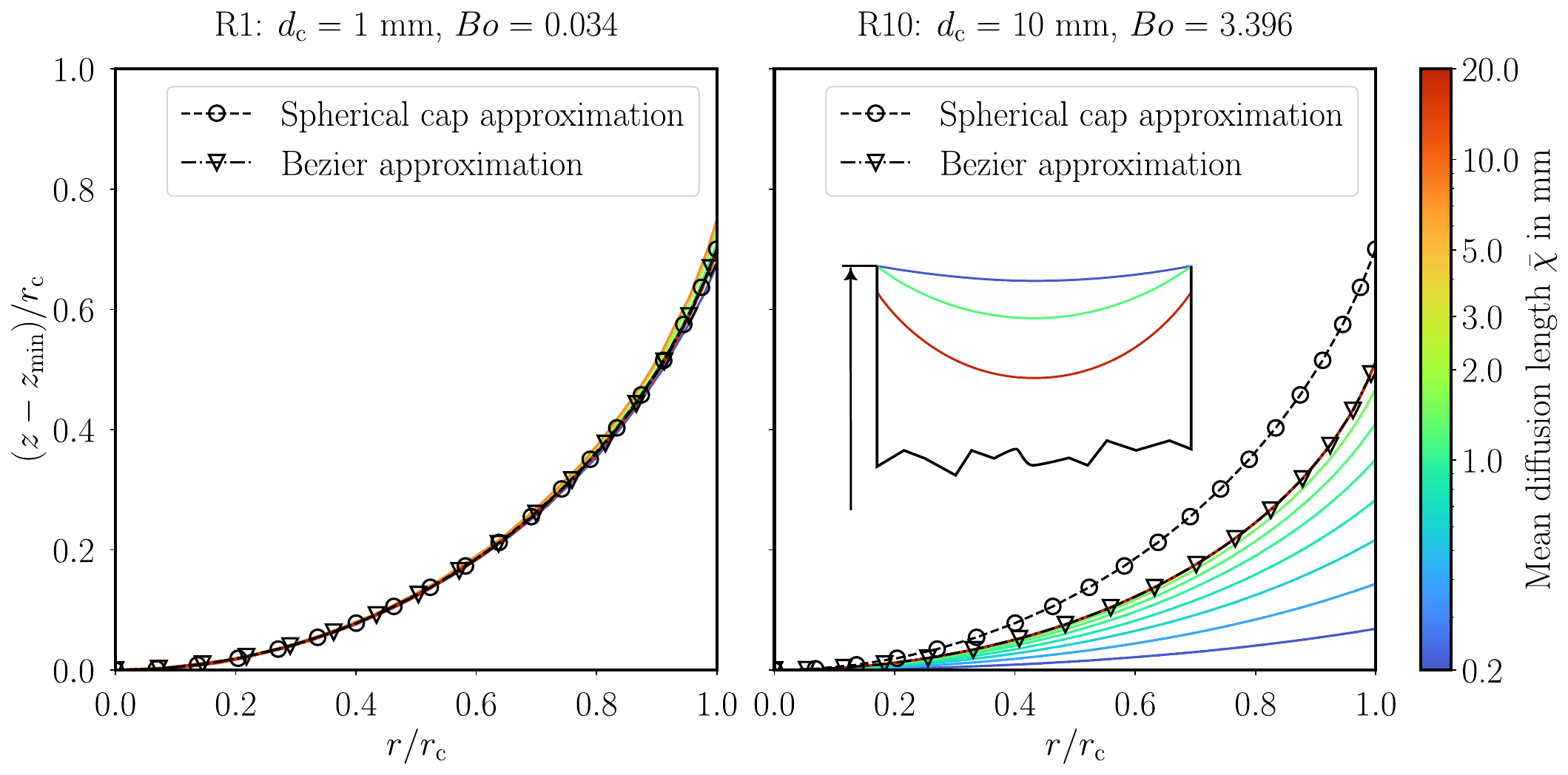}}%
    \put(0.53157669,0.24682831){\color[rgb]{0,0,0}\rotatebox{90}{\makebox(0,0)[t]{\lineheight{1.25}\smash{\begin{tabular}[t]{c}$h_{\mathrm{c}}$\end{tabular}}}}}%
    \put(0,0){\includegraphics[width=\unitlength,page=2]{Figure_7.pdf}}%
    \put(0.81140135,0.29523281){\color[rgb]{0,0,0}\makebox(0,0)[t]{\lineheight{0}\smash{\begin{tabular}[t]{c}$\bar{\chi}$\end{tabular}}}}%
    \put(0,0){\includegraphics[width=\unitlength,page=3]{Figure_7.pdf}}%
    \put(0.10852353,0.23588984){\color[rgb]{0,0,0}\rotatebox{90}{\makebox(0,0)[t]{\lineheight{1.25}\smash{\begin{tabular}[t]{c}$h_{\mathrm{c}}$\end{tabular}}}}}%
    \put(0.2710808,0.26203914){\color[rgb]{0,0,0}\makebox(0,0)[t]{\lineheight{0}\smash{\begin{tabular}[t]{c}$\bar{\chi}$\end{tabular}}}}%
    \put(0,0){\includegraphics[width=\unitlength,page=4]{Figure_7.pdf}}%
  \end{picture}%
\endgroup%

            \color{black}
            \caption[Meniscus sections of round capillaries.]{\revone{Meniscus sections of the round capillaries with $d\ped{c}=1\unit{mm}$ (left) and $d\ped{c}=10\unit{mm}$ (right) for various mean diffusion lengths $\bar{\chi}$ are displayed in solid colored lines. Utilizing symmetry, only the right arc of the section is depicted and the sections are shifted vertically to align their deepest points. Black lines show the spherical cap \cite{grunding_capillary_2019} and the Bezier approximation \cite{lewis_calculation_2022}. The upper capillary boundary constrains the meniscus shape, especially in the R10 case at smaller mean diffusion lengths which results in a different meniscus shape. This effect is less pronounced in the R1 case due to the larger mean diffusion length relative to the capillary diameter.}}
            \label{fig:roundCapillary_meniscusShape}
        \end{figure}

        Looking at the shape of the menisci in \cref{fig:SEresultsR1} it is noticeable that the fluid surface shapes are similar for all displayed mean diffusion lengths. In order to better compare the shapes, the half sections for the R1 and R10 case are visualized in \cref{fig:roundCapillary_meniscusShape}.
        
        % For smaller mean diffusion lengths, the capillary opening constrains the meniscus shape, especially in the R10 case, resulting in a larger contact angle at the wall and thus a different meniscus shape. This constraint is less significant in the R1 case, where the meniscus retains a similar shape for the same mean diffusion length. 

        \cref{fig:roundCapillary_meniscusShape} shows that the meniscus deepest point lies in the capillary center ($r/r\ped{c}=0$) in all simulated cases. The gradient $\diff z/\diff r\xrightarrow{}0$ for $r\xrightarrow{}0$ yielding a continuously differentiable surface for $r\in (-r\ped{c},r\ped{c})$. Further, the water rises at the capillary walls ($r/r\ped{c}=1$) in both cases in order to meet the contact angle boundary condition. \color{\revonecolor}In the R1 case, the apparent contact angle at the wall matches the specified contact angle (here $\theta=20^{\circ}$) well for all investigated values of $\bar{\chi}$. However, in the R10 case for $\bar{\chi}<3\unit{mm}$, the apparent contact angle is greater than the specified contact angle. In round capillaries of constant width and without an upper boundary, the liquid surface shape does not depend on liquid filling height or mean diffusion length respectively. However, the investigated capillaries have an upper boundary which leads to the apparent contact angle decreasing at the beginning (starting from a full capillary) until the apparent contact angle matches the specified contact angle.\color{black} % Since the investigated capillaries have an upper limit, the contact angle increases in order to still conserve the liquid volume according to \cref{eq:volumeConservation}. That leads to the contact angle decreasing at the beginning (starting from a full capillary) until the specified contact angle is reached. % This behavior occurs if the meniscus elevation through curvature is higher than the mean diffusion length $\bar{\chi}$.
        
        %The meniscus shape where this boundary condition is satisfied, is referred to as \textit{fully developed shape} in the following. This shape corresponds to the meniscus of a motionless fluid in a round capillary of infinite height. In the R10 case, the meniscus is not fully developed for $\bar{\chi}<3\unit{mm}$, as visible in the right graph. In those simulations, the water surface reaches the top of the capillary which results in an increasing contact angle in order to comply with the volume conservation. However, in the R1 case, the meniscus is fully developed for all mean diffusion lengths due to the liquid elevation not being limited at the top of the capillary.

        Differences in meniscus shape depend strongly on the Bond number $Bo$ which relates gravitational forces to surface tension forces. It is defined as $Bo=(\rho\ped{l}-\rho\ped{g}) gr\ped{c}^2/\sigma$ for round capillaries where $g$ is the gravity constant. For capillaries with a sufficiently small diameter where $Bo<1$, surface tension dominates the force balance at the interface which results in an approximate shape of a spherical cap \cite{lewis_calculation_2022, grunding_capillary_2019}. The surface formation simulation of the R1 case on the left side of \cref{fig:roundCapillary_meniscusShape} shows only slight deviations from the spherical cap approximation.

        However, for large diameters with $Bo>1$, the spherical cap approximation fails to estimate the meniscus shape as shown in \cref{fig:roundCapillary_meniscusShape} for the R10 case. This is due to the dominance of gravitational forces, which result in a flattened liquid surface at the center of the capillary \cite{lewis_calculation_2022}. Using the well-known Young-Laplace equation, an equation for the meniscus shape can be derived which also takes gravity effects into consideration. Lewis and Matsuura \cite{lewis_calculation_2022} proposed a method to approximate this equation using a Bezier curve approach.

        Using the Bezier curve method, the meniscus shape in both cases is solved numerically with Python 3.10 and the SciPy 1.13.1 package \cite{virtanen_scipy_2020}. Therefore, the \textit{trust-constr} minimization algorithm \cite{nocedal_numerical_2006,lalee_implementation_1998} is used to approximate the modified Young-Laplace equation and incorporate constraints, initial and boundary conditions. For both capillaries, the fully developed menisci in SE coincide with the Bezier approximation. Thus, SE accomplishes to approximate the Young-Laplace equation for fully developed menisci.

        The R1 and R10 cases are used to validate the presented method and can be compared to the analytical solution from Stefan \cite{stefan_uber_1871}. According to Stefan, the evaporation rate of round capillaries is calculated by
        ~
        \begin{equation}
            \label{eq:stefanEvapRate}
            \dot{m}\ped{ev}=A\ped{c}D\frac{\rho\ped{g}(Y\ped{v,sat}-Y\ped{v})}{\bar{\chi}}.
        \end{equation}

        The analytical solution from Stefan does not consider an initial mass boundary layer over the capillary which leads to $\dot{m}\ped{ev}\xrightarrow{}\infty$ for $\bar{\chi}\xrightarrow{}0$. To take the initial mass boundary layer above the capillary into account, Camassel et al. \cite{camassel_evaporation_2005} use a modified diffusion length $\bar{\chi}^*=\bar{\chi}+\delta$ where $\delta$ is the assumed uniform thickness of the external mass boundary layer. In order to better compare the analytical solution with the experimental results, the boundary layer thickness $\delta$ is calculated from the experimental data using the following equation by Camassel et al. \cite{camassel_evaporation_2005}
        ~
        \begin{equation}
            \delta=\frac{\rho\ped{g}D(Y\ped{v,sat}-Y\ped{v})}{\dot{m}\ped{ev}(\bar{\chi}=0)A\ped{c}}.
        \end{equation}

\begin{figure}[tb]
            \centering
            \includegraphics[width=\linewidth]{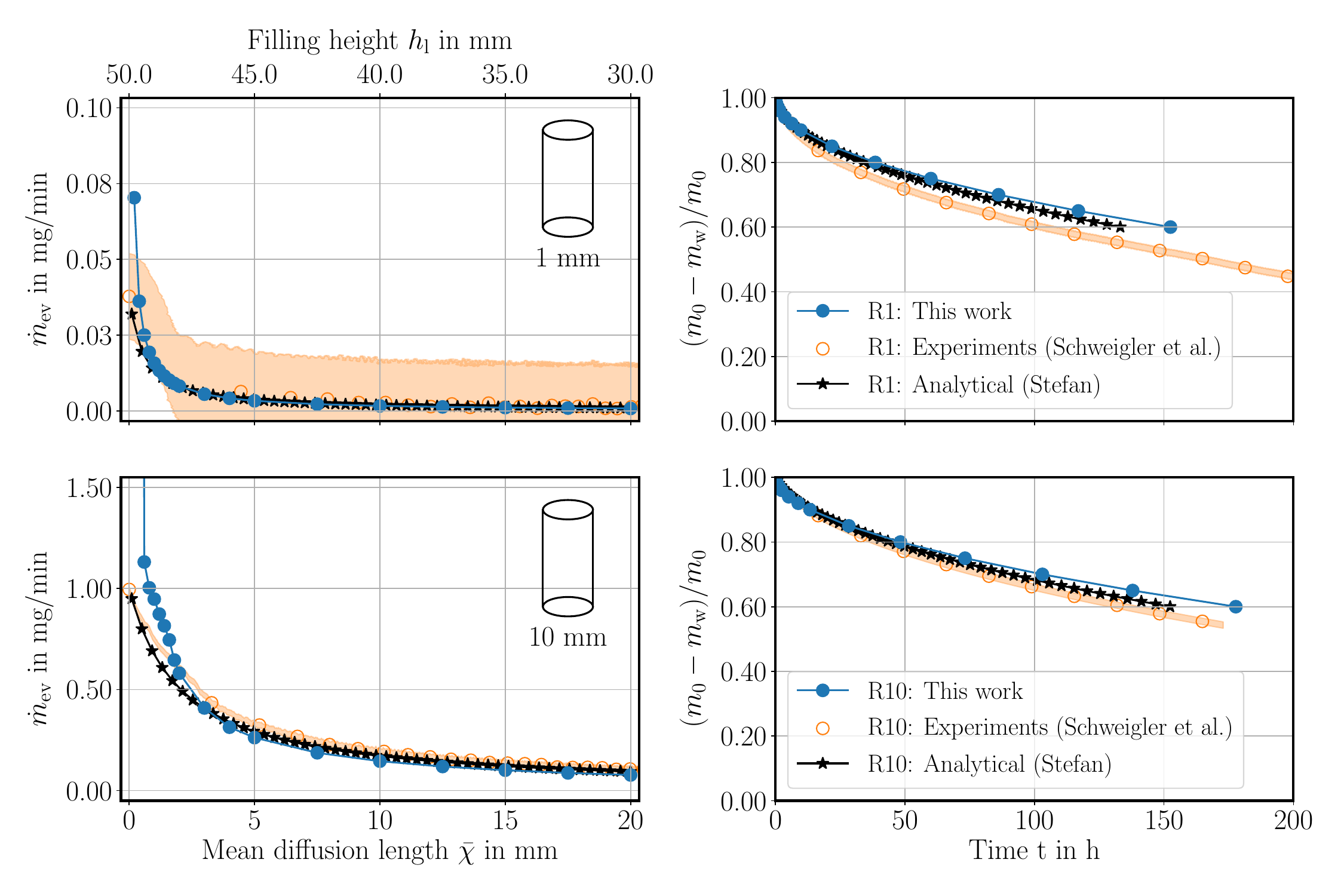}
            \caption[Evaporation rate of round capillaries.]{\revtwo{Evaporation rate (left) and normalized water mass (right) for round capillaries using the proposed method in this work, the experimental results from Schweigler et al. \cite{schweigler_analyse_2017} and the analytical solution in \cref{eq:stefanEvapRate} with the modified diffusion length $\bar{\chi}^*$. For the experiments, the band around the data points shows the scale measurement error.}}
            \label{fig:Results_RoundCapillaries}
        \end{figure}

        The simulated results are compared with the experiments and the analytical solution in \cref{fig:Results_RoundCapillaries}.
    
        The left graphs in \cref{fig:Results_RoundCapillaries} show the evaporation rate for the 1\unit{mm} and the 10\unit{mm} capillary. In both cases, the evaporation rate is highest for $\bar{\chi}=0.2\unit{mm}$ and decreases at higher mean diffusion lengths. In the R1 case, the simulated evaporation rate lies perfectly within the experimental error band for $\bar{\chi}\geq0.4\unit{mm}$ and further matches the analytical solution for $\bar{\chi}\geq1\unit{mm}$. The simulated evaporation rate in the R10 case deviates more from the experiments in the beginning compared to the R1 case, but shows good agreement for $\bar{\chi}>2\unit{mm}$ nonetheless. Deviations in the beginning must not be caused by simulation errors but could also be attributed to small changes in the meniscus position at which the logging started in the experiments.
        
        Although the proposed method provides reasonable results in most simulations, the evaporation rate for $\bar{\chi}=0.2\unit{mm}$ in the R10 case is exceptionally high. This can be traced back to a failed meshing process in sHM which occurs irregularly and is part of further improvement.

        The simulations and the analytical solution in both cases predict a slower decrease of water mass on the long run compared to the experiments as depicted in the right graphs of \cref{fig:Results_RoundCapillaries}. This can be an indicator for additional effects that promote evaporation which are not considered in the analytical solution and the simulation. Further work should investigate the influence of buoyancy-driven convection on the evaporation rate, since temperature gradients in the experimental setup could induce currents that enhance the transport of water vapor out of the capillary.
        
    \subsection{Square capillaries}
        \label{sec:ResultsSquareCapillaries}
\begin{figure}[tb]
            \centering
            \includegraphics[width=0.07\linewidth]{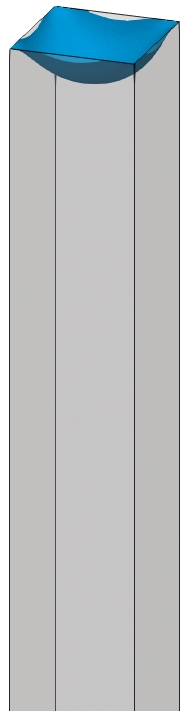}
            \includegraphics[width=0.07\linewidth]{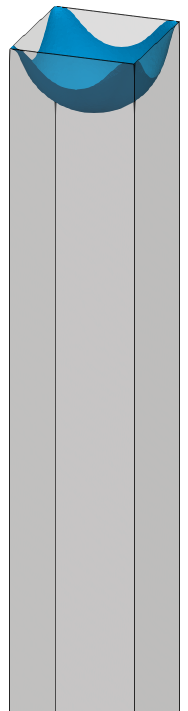}
            \includegraphics[width=0.07\linewidth]{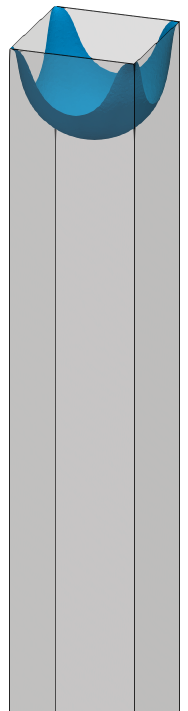}
            \includegraphics[width=0.07\linewidth]{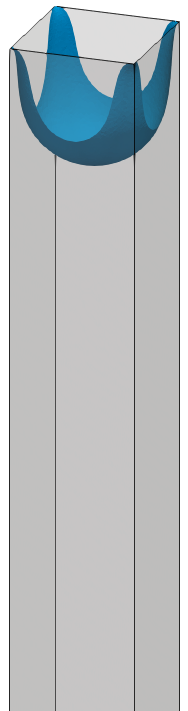}
            \includegraphics[width=0.07\linewidth]{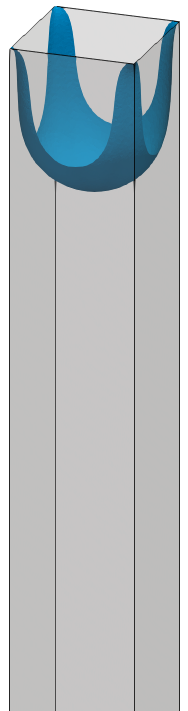}
            \includegraphics[width=0.07\linewidth]{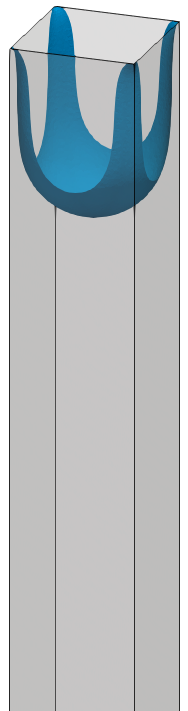}
            \includegraphics[width=0.07\linewidth]{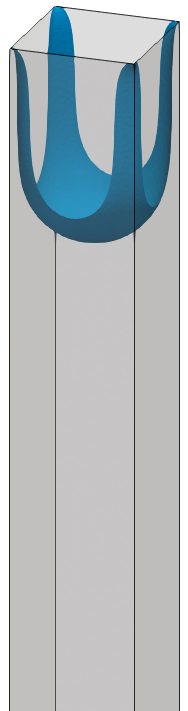}
            \includegraphics[width=0.07\linewidth]{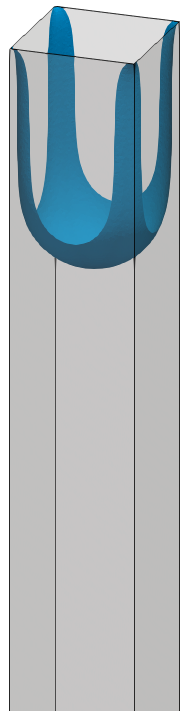}
            \includegraphics[width=0.07\linewidth]{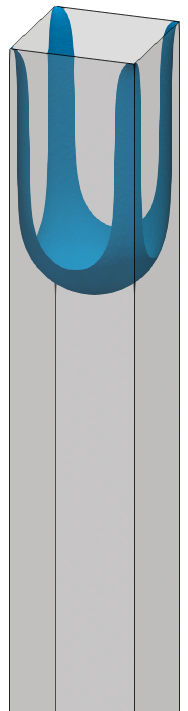}
            \includegraphics[width=0.07\linewidth]{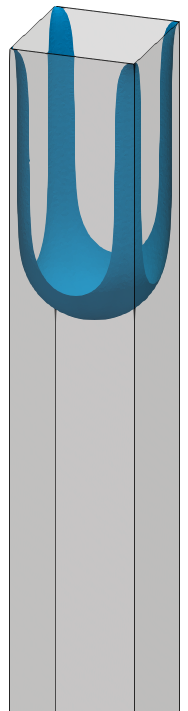}
            \includegraphics[width=0.07\linewidth]{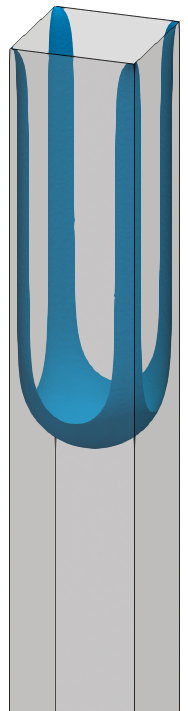}
            \includegraphics[width=0.07\linewidth]{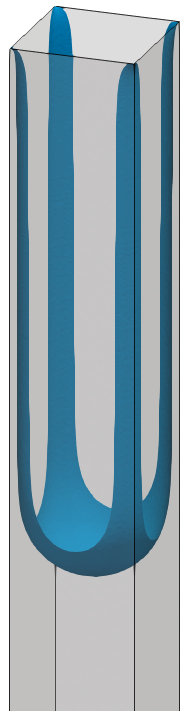}
            \includegraphics[width=0.07\linewidth]{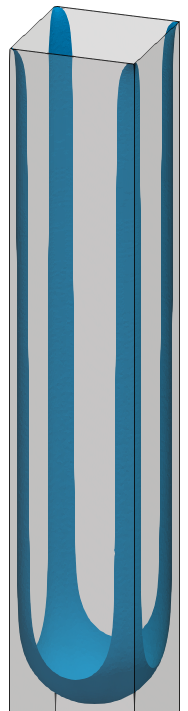}
            \par\vspace{2mm}
            \fontsize{12pt}{12pt}\selectfont
            \def\svgwidth{\linewidth}
            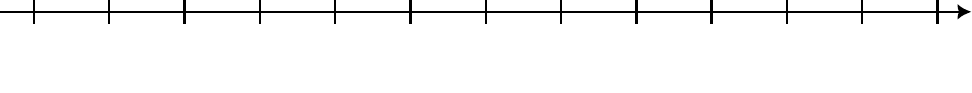
            \caption[Fluid surface of square capillary as simulated in Surface Evolver.]{Simulated fluid surface from SE for the square capillary with a width of $a\ped{c}=1\unit{mm}$, height of $h\ped{c}=50\unit{mm}$ and a contact angle of $\theta=20\,^{\circ}$. For higher mean diffusion lengths $\bar{\chi}$, liquid fingers form in the corners of the capillary and rise to the top of the capillary.}
            \label{fig:SEresultsS1}
        \end{figure}

        Previous experiments by Schweigler et al. \cite{schweigler_evaporation_2018} show that the evaporation rate increases significantly for capillaries with a square cross-section compared to round capillaries with a similar cross-sectional area. The reason for higher evaporation rates is the presence of liquid fingers in the corners of the capillary, resulting in a local reduction of the diffusion length \cite{schweigler_evaporation_2018}. In order to elaborate the influence of liquid fingers on the evaporation rate, simulations of square capillaries with an inner edge length of $1\unit{mm},\,4\unit{mm}\text{ and }13\unit{mm}$ (cases S1, S4, S13) are performed. As a first step, the liquid surface formation is simulated in SE for the mean diffusion lengths in \cref{eq:meanDiffusionLengths}. \cref{appSec:SEIterationProcess} provides an example of the iteration process from a flat plane to a curved meniscus. The resulting water surface for the case S1 is shown as an example in \cref{fig:SEresultsS1}.

        The water surfaces in \cref{fig:SEresultsS1} are axisymmetric in relation to the capillary principal axis. For the specified geometry, the contact angle of $\theta=20\,^{\circ}$ does not satisfy Concus and Finn's condition for a bounded solution of the Young-Laplace equation \cite{concus_behavior_1969} which results in liquid fingers, rising inside the corners of the capillary. Based on investigations from Chauvet et al. \cite{chauvet_depinning_2010}, the liquid finger length strongly depends on the corner roundedness of the capillary. The investigated square capillary has sharp corners which in theory results in infinitely high finger lengths. Since the capillary height is limited in the simulation, the liquid fingers stop at the top of the capillary as depicted in \cref{fig:SEresultsS1}.

\begin{figure}[H]
            \centering
            \includegraphics[width=0.95\linewidth]{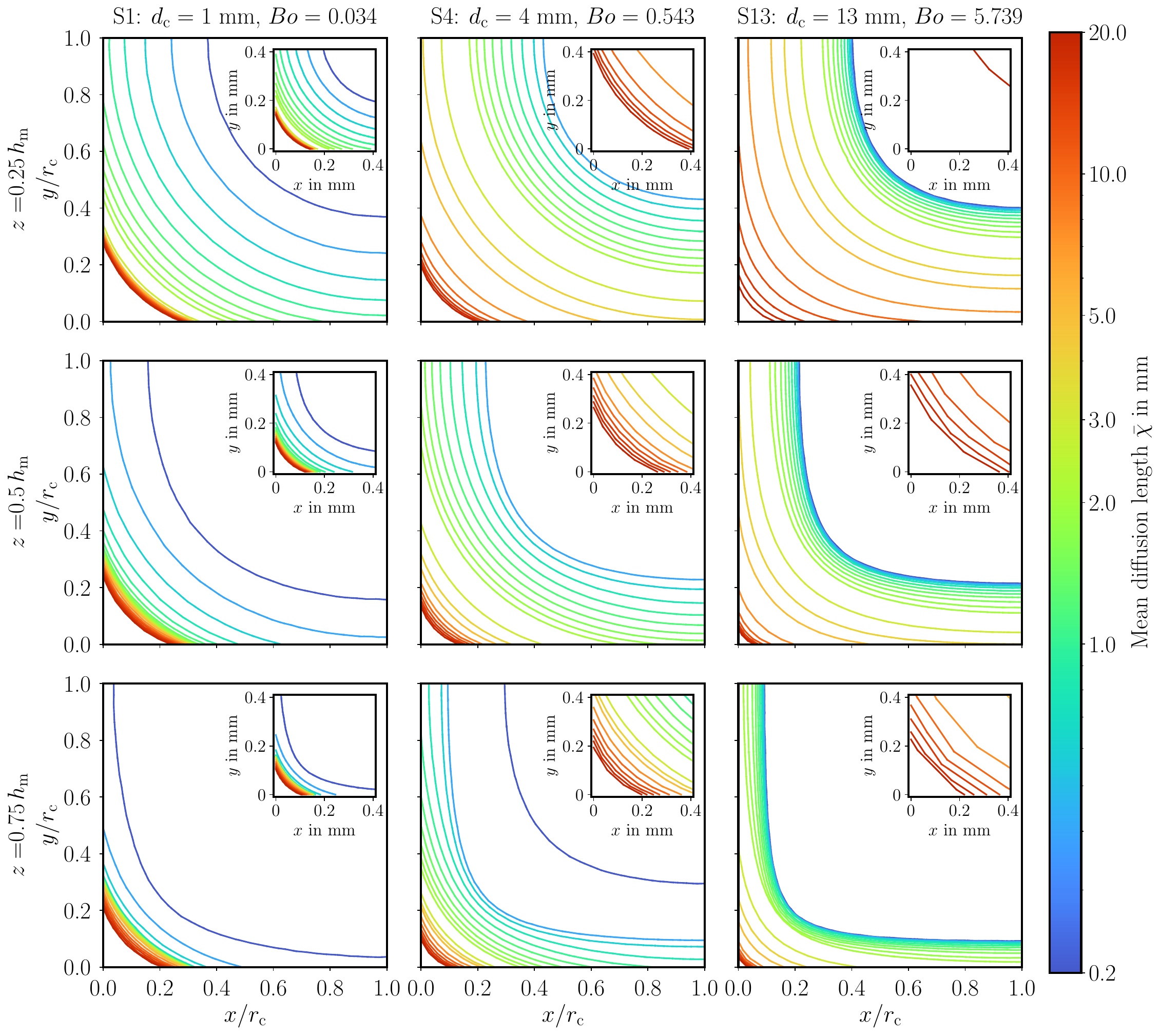}
            \caption[Sections of liquid fingers at different heights.]{Section of a single liquid finger in the S1, S4 and S13 case for various mean diffusion lengths at 25 \%, 50 \% and 75 \% of the total meniscus height $h\ped{m}$. $x$ and $y$ coordinates are normalized by the half capillary width $r\ped{c}$. Inset axes show absolute values in the proximity of the corner.}
            \label{fig:fingersShape}
        \end{figure}

        \cref{fig:fingersShape} shows the cross-section of a single liquid finger for the S1, S4 and S13 case and various mean diffusion lengths. The liquid fingers are each cut at 25 \%, 50 \% and 75 \% of the total meniscus height $h\ped{m}$ which is the distance between the surfaces deepest and highest points. 
        
        In order to comply with the contact angle boundary condition, the fingers are horizontally curved. For higher mean diffusion lengths, the fingers rise vertically in the corners and their thickness approaches a minimum value. Although the capillary widths between the S1 and S13 cases are an order of magnitude apart, the minimum finger thickness is in the same order of magnitude. Furthermore, the minimum finger thickness does not change significantly over the finger height. In the S1 case, the finger thickness quickly approaches a limit for small mean diffusion lengths, whereas in the S13 case, the fingers are developed at much higher mean diffusion lengths. \cref{fig:fingersShape} also shows that the water surface in the S13 case tends to represent the outer square shape more strongly than the S1 case. This is attributed to the Bond number being higher, thus gravitational effects dominate capillary effects.

\begin{figure}[H]
            \centering
            \includegraphics[width=0.95\linewidth]{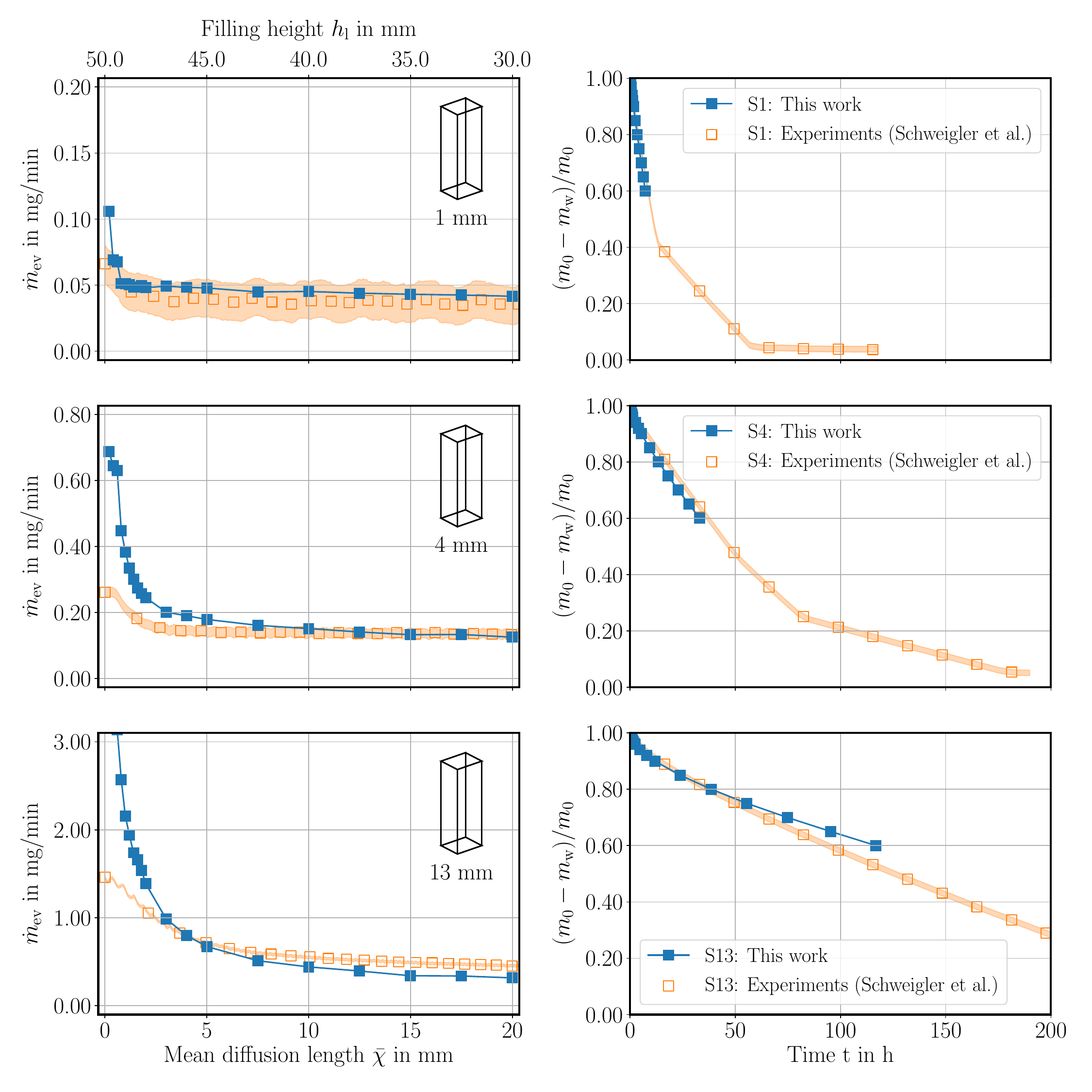}
            \caption[Evaporation rate of square capillaries.]{\revtwo{Evaporation rate (left) and normalized water mass (right) for square capillaries using the proposed method in this work and the experimental results from Schweigler et al. \cite{schweigler_analyse_2017}. For the experiments, the band around the data points shows the measurement error.}}
            \label{fig:Results_SquareCapillaries}
        \end{figure}

        After simulating the water surface, the evaporation rate is determined according to \cref{sec:diffusionSim}. The simulative results and the experimental values are shown in \cref{fig:Results_SquareCapillaries}.

        The total evaporation rate in the S1 case on the upper left of \cref{fig:Results_SquareCapillaries} matches the experimental data well, although the simulated evaporation rates are slightly higher than the mean values from the experiment. However, the general trend of the experiment is well replicated.

        In the S1 case, the measured water mass can be separated into three distinct phases that were previously described by Chauvet et al. \cite{chauvet_three_2009} (see \cref{sec:introduction}). The transition from CRP to FRP is characterized by a reduction in evaporation rate, attributed to the depinning of liquid fingers from the open end of the capillary. Given that the simulations are performed with sharp capillary corners, the natural depinning that occurs due to rounded corners cannot be replicated in the simulations.

        In the S4 and S13 case, the simulation also overestimates the evaporation rate in the beginning but approaches the experimental values for increasing mean diffusion lengths. The deviations in the beginning only have a minor influence on the predicted water mass, as depicted in the right graphs. However, the evaporation rate in the S13 case is underestimated for $\bar{\chi}\geq5\unit{mm}$ which leads to a slower drying, compared to the experiment, as visualized in the lower right graph.

        The previous observations show that the proposed method enables the determination of evaporation rate also for square capillaries. Expected liquid fingers could be verified in the simulations which lead to a higher evaporation rate compared to round capillaries. Due to the anti-proportional relation between time and evaporation rate (see \cref{eq:timeDependence}), minor deviations in evaporation rate result in slower drying compared to the experiment. This behavior is more pronounced in case of wide capillaries and can also be observed for the round capillaries which might be attributed to large scale effects, such as convection inside the capillary.

    \subsection{Temperature dependence}
        \label{sec:ResultsTempDependence}
        Previous simulations were only conducted for a temperature of $T_0=23\,^{\circ}$C and a relative humidity of $\varphi_0=17\,\%$. In order to predict evaporation rates for different thermodynamical boundary conditions, only the vapor diffusion simulations are repeated for various temperatures while the relative humidity is kept constant. Simulations are conducted as an example for the S1 case from \cref{sec:ResultsSquareCapillaries}. Thermal influences on contact angle and surface tension are neglected, thus the same set of surfaces from SE is used for all investigated temperatures.
        
\begin{figure}[tb]
            \centering
            \includegraphics[width=\linewidth]{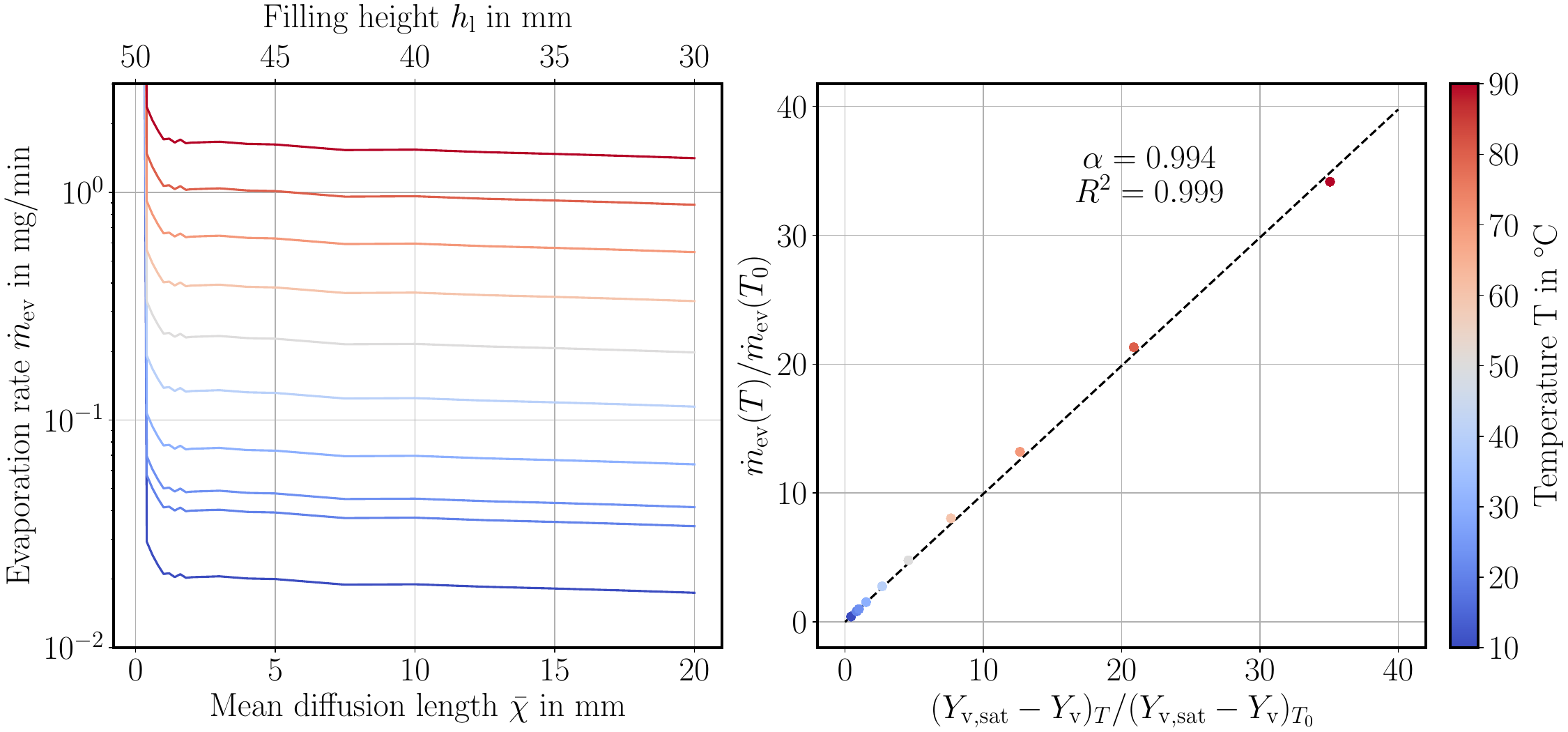}
            \caption{Evaporation rate (left) in a square capillary with $a\ped{c}=1\unit{mm},\,h\ped{c}=50\unit{mm},\,\theta=20^{\circ}$ for temperatures ranging from $10^{\circ}$C up to $90\,^{\circ}$C. The right plot shows the evaporation rate at the temperature $T$ normalized by the evaporation rate at $T_0=23\,^{\circ}$C for specified mass fraction ratios.}
            \label{fig:Results_tempDependence}
        \end{figure}

        Since the temperature is not a direct input to the diffusion simulation, it must be used to determine the vapor mass fractions at the meniscus and environment, according to \cref{eq:saturationPressure,eq:moleFraction,eq:massFraction}. In order to evaluate the evaporation rate, the gas mixture density $\rho\ped{g}$ and the diffusion coefficient $D$ in \cref{eq:integratedMassFraction} must be adapted. The calculated values are listed in \cref{tab:tempDependenceParams} and the respective results are plotted in \cref{fig:Results_tempDependence}.
    
        \begin{table}[tb]
            \centering
            \begin{tabular}{lcccccccccc}
                \hline
                $T$ in $^{\circ}$C & 10 & 20 & 23 & 30 & 40 & 50 & 60 & 70 & 80 & 90 \\
                $\varphi$ in \% & 17 & 17 & 17 & 17 & 17 & 17 & 17 & 17 & 17 & 17 \\
                \hline
                $Y\ped{v}$ in g/kg & 1.30 & 2.48 & 2.98 & 4.51 & 7.86 & 13.17& 21.36 & 33.65 & 51.69 & 77.72 \\
                $Y\ped{v,sat}$ in g/kg & 7.68 & 14.71 & 17.71 & 26.90 & 47.32 & 80.64 & 134.2 & 219.9 & 359.1 & 594.2 \\
                $\rho\ped{g}$ in kg/m$^3$ & 1.24 & 1.20 & 1.18 & 1.15 & 1.11 & 1.06 & 1.01 & 0.96 & 0.89 & 0.81 \\
                $D$ in mm$^2$/s & 22.88 & 24.31 & 24.75 & 25.78 & 27.28 & 28.83 & 30.41 & 32.02 & 33.67 & 35.36 \\
                \hline
            \end{tabular}
            \caption{Thermodynamical properties of the air-vapor mixture for temperatures between 10 $^{\circ}$C and 90 $^{\circ}$C. Densities are calculated using the ideal gas law and diffusion coefficients are determined using the method by Fuller et al. \cite{fuller_diffusion_1969} according to \cite{vdi_e_v_vdi-warmeatlas_2013}.}
            \label{tab:tempDependenceParams}
        \end{table}
    
        The results in \cref{fig:Results_tempDependence} on the left side show that the evaporation rate increases significantly with higher temperatures and that the graphs are self-similar to each other because the water surface has not been adjusted to account for temperature changes. In the logarithmic scale of the plot, the evaporation rate increases almost linear which allows the assumption of an exponential relation $\dot{m}\ped{ev}\propto \exp(T)$. In fact, the evaporation rate depends on the mass fraction gradient $\nabla Y$ which is related to the temperature in terms of the Antoine \cref{eq:saturationPressure}.
        
        To obtain a general correlation for $\dot{m}\ped{ev}(T)$, the influence of the mass fraction difference $Y\ped{v,sat}-Y\ped{v}$ on the evaporation rate is investigated. Therefore, the relationship
        ~
        \begin{equation}
            \label{eq:evapRateTempDependence}
            \frac{\dot{m}\ped{ev}(T)}{\dot{m}\ped{ev}(T_0)} = \alpha\frac{(Y\ped{v,sat}-Y\ped{v})_T}{(Y\ped{v,sat}-Y\ped{v})_{T_0}}
        \end{equation}
    
        is proposed, with the proportionality factor $\alpha$. The proportionality factor is determined by normalizing all evaporation rates in \cref{fig:Results_tempDependence} using the evaporation rate at $T_0=23\,^{\circ}$C. Further, the vapor mass fraction differences at each temperature are normalized by the vapor mass fraction difference at $T_0=23^{\circ}$C. The resulting plot is depicted in \cref{fig:Results_tempDependence} on the right side.
    
        The plot shows that the normalized evaporation rate strongly correlates with the vapor mass fraction ratio. The correlation factor is $R^2=0.99$ and the proportionality factor evaluates to $\alpha=0.99$. According to the simulative results the evaporation rate can be easily adapted to various temperatures, depending on the vapor mass fraction difference. This allows to simulate the evaporation rate only for one specific temperature and calculate the evaporation rate for a different temperature using \cref{eq:evapRateTempDependence}. Nevertheless, experiments with higher temperatures must be performed in the future, in order to validate the presented results and provide information, whether additional non-linear effects must be included in the simulation.

    \subsection{Runtime}
        \label{sec:runtime}
        The major advantage of the presented method is the short simulation runtime. All simulations were carried out on a high performance cluster that runs Linux and uses \textit{Intel Xeon Gold 6342} CPUs (Central Processing Unit). The simulation process is subdivided into four steps in order to set the computational resources according to the individual computational requirements. The respective steps and resources are listed in \cref{tab:computationalResources}.

        \begin{table}[tb]
            \centering
            \begin{tabular}{cccc}
                \hline
                Step    & Description                   & No. of cores  & Max. memory per core in GB \\
                \hline
                1       & Surface formation simulation  & 1             & 0.5\\
                2       & Meshing preparation           & 1             & 16\\
                3       & Meshing (sHM)       & 4             & 8\\
                4       & Vapor diffusion simulation    & 1             & 16\\
                \hline
            \end{tabular}
            \caption{Computational resources for the presented simulations in \cref{sec:ResultsRoundCapillaries,sec:ResultsSquareCapillaries,sec:ResultsTempDependence}.}
            \label{tab:computationalResources}
        \end{table}

\begin{figure}[tb]
            \centering
            \includegraphics[width=\linewidth]{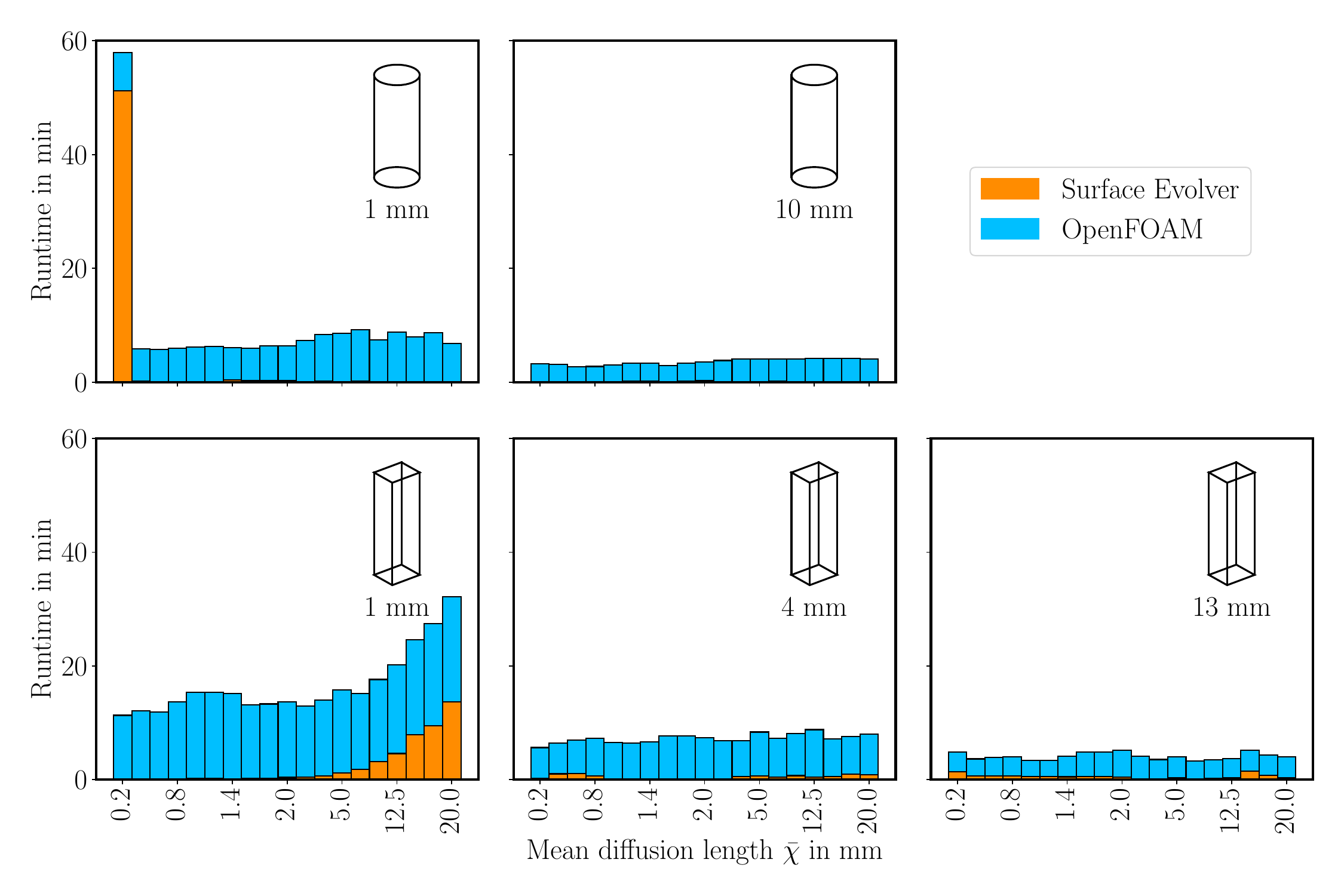}
            \caption{Absolute runtime for the cases R1, R10, S1, S4 and S13 from \cref{sec:ResultsRoundCapillaries,sec:ResultsSquareCapillaries} split up into SE runtime and OF runtime.}
            \label{fig:runtimes}
        \end{figure}

        In the following, the first step is considered separately, since it only involves SE calculations while step 2,3 and 4 incorporate Linux commands and OF calculations. The runtime for all simulations from \cref{sec:ResultsRoundCapillaries,sec:ResultsSquareCapillaries} are visualized in \cref{fig:runtimes}.
        
        With the presented method we are now able to run simulations in 5~min to 60~min to cover up to 178~h of physical time for round capillaries. The simulation runtime is dominated by the calculations in OF while the SE runtime is negligible. Only in one case, the SE struggles to simulate the water surface for $\bar{\chi}=0.2\unit{mm}$ (see \cref{fig:runtimes} upper left) which is attributed to slow surface energy decrease. This behavior is likely caused by bad initial conditions for high filling heights as discussed in \cref{appSec:reliability} and is part of further investigations.
        
        The square capillary simulation runtime is between 5~min and 30~min and is also dominated by the OF simulations, although the SE runtime increases continuously for higher mean diffusion lengths in the S1 case. This is related to the formation of liquid fingers and their increasing height.

        Although the simulated physical time ranges from $t\ped{phys}\approx 8\unit{h}$ in the S1 case to ${t\ped{phys}\approx 178\unit{h}}$ in the R10 case, the maximum runtime of the simulations is only ${t\ped{run,max}\approx 1\unit{h}}$. Since the simulations for various mean diffusion lengths $\bar{\chi}$ are independent from each other, it must be emphasized that all simulations can run at the same time if sufficient CPUs are provided. To the best of our knowledge, there is no other method with a comparable runtime while still providing accurate results. 

    \subsection{Generalization}
\begin{figure}[tb]
            \centering
            \includegraphics[width=\linewidth]{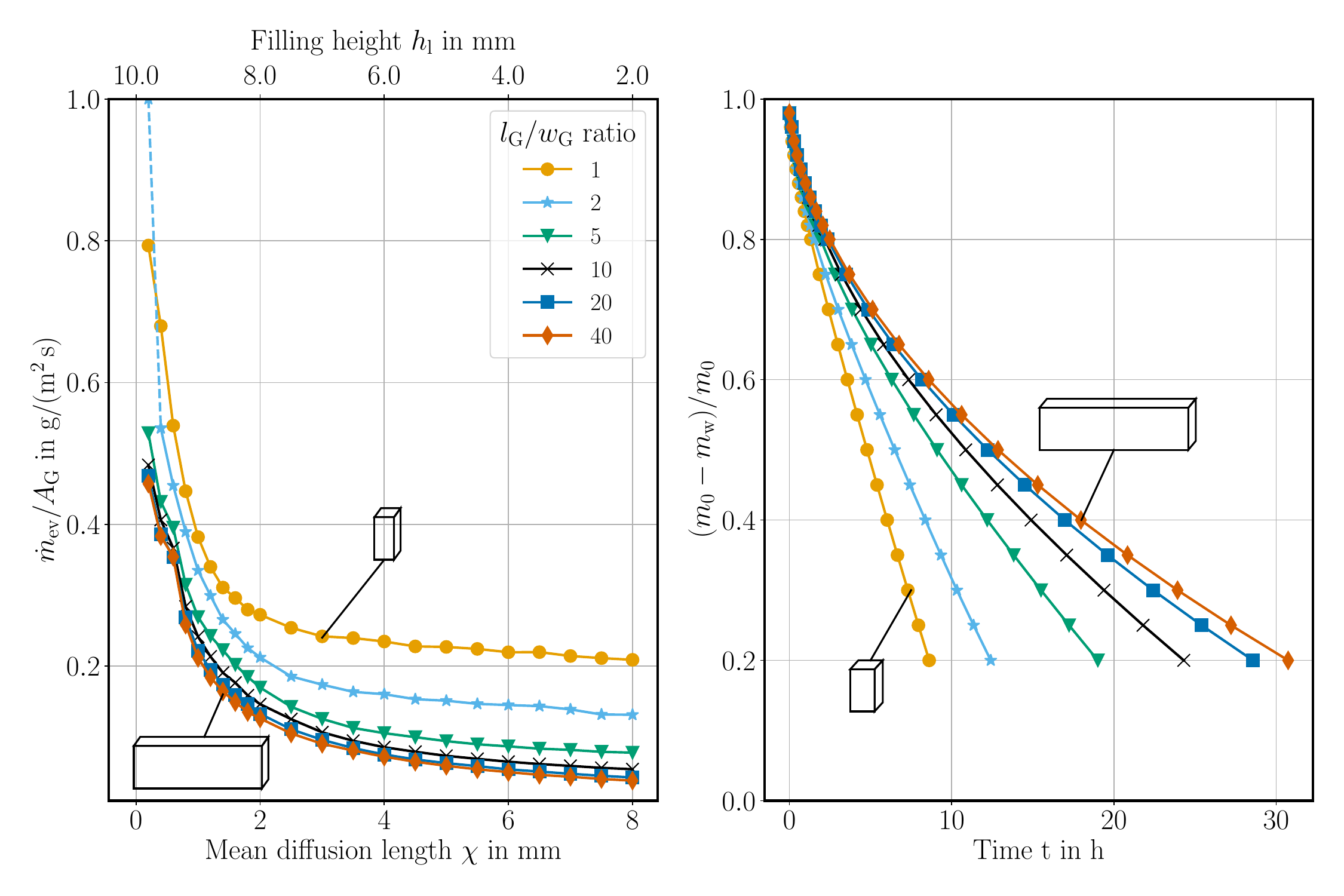}
            \caption{Cross-section normalized evaporation rate (left) of water in humid air and normalized total water mass (right) for a gap with various length-to-width ratios at $T=23\,\unit{^{\circ}C},\, p=101325\unit{Pa},\, \varphi=17\,\%$. The dashed line in the $l\ped{G}/w\ped{G}=2$ case indicates an outlier due to a meshing error in sHM.}
            \label{fig:Results_Gap}
        \end{figure}

        Idealized geometries like round or square capillaries are rarely found in real-world applications. Nevertheless, rectangular gaps are often created by ribs that are used to increase stiffness of industrial housings. In order to test the proposed method for a wider range of applications, simulations are conducted for flat gaps with various length-to-width ratios. The gap width and height are kept constant at $w\ped{G}=3\unit{mm}$ and $h\ped{G}=10\unit{mm}$ respectively. The gap length is set according to a length-to-width ratio ranging from $1\leq l\ped{G}/w\ped{G}\leq 40$. The thermodynamical boundary conditions are $T=23\,\unit{^{\circ}C},\, p=101325\unit{Pa},\, \varphi=17\,\%$. Using the presented method, the simulated evaporation rate and mass decrease over time are visualized in \cref{fig:Results_Gap}. 
        
        To better compare the evaporation rate in various gaps, the results are normalized by the cross-sectional area $A\ped{G}=l\ped{G}w\ped{G}$ of the gap. The graph clearly shows that the normalized evaporation rate is highest for $l\ped{G}/w\ped{G}=1$ and decreases towards a limit for higher length-to-width ratios. As described in \cref{sec:ResultsSquareCapillaries}, the liquid finger thickness increases only slightly for higher capillary widths which results in a decreasing influence of liquid fingers for higher $l\ped{G}/w\ped{G}$ ratios. The mass plot allows for a similar conclusion, as the water volume and absolute evaporation rate depend linearly on the gap cross-sectional area.
        With the presented method, the evaporation in arbitrary capillaries can be simulated within a reasonable simulation time, as far as the geometry can be described in SE. Future work should focus on incorporating STL (stereolithography) files in SE and conducting experiments to validate the simulated results shown here.
    
    % Conclusion
    \section{Conclusion}
\label{sec:conclusion}
This work presents a novel semi-transient simulation approach to estimate evaporation rates in small gaps, relevant to industrial applications. By decoupling the simulation of the liquid-gas interface from the vapor diffusion simulation, the approach addresses the challenges of long simulation times and numerical errors associated with traditional methods like Volume-of-Fluid. The proposed method utilizes SE to simulate the fluid surface formation, considering capillary effects and then employs OF to simulate vapor diffusion in the gas domain. The source code is uploaded to the Bosch Research GitHub \footnote{\url{https://github.com/boschresearch/sepMultiphaseFoam/tree/publications/novelSimulationApproachForEvaporationProcesses}}. The decoupling allows for the evaporation rate to be calculated independently of previous time steps, enabling highly parallelized simulations. 

The key advantage of this method is its speed. Simulations for round, square and rectangular capillaries were conducted, covering physical times up to 178 hours, that are completed within 1 hour of computational time. The simulations significantly outperform conventional methods like Volume-of-Fluid, which require small time steps to resolve the interface motion coupled with evaporation, resulting in prohibitively long simulation times for realistic scenarios.

The numerical investigations conducted for round and square capillaries demonstrate good agreement with experimental results for a wide range of capillary sizes. While the approach effectively predicts liquid finger formation in square and rectangular capillaries and captures the associated increase in evaporation rate, some limitations were observed. In narrow capillaries with large filling heights, the accurate prediction of the liquid surface and the evaporation rate becomes more challenging, leading to deviations from experimental values. Simulations were also conducted for gaps, showing the wide applicability of this method. The simulation runtime remains significantly lower than fully coupled approaches, highlighting the computational efficiency of the proposed method.

Future work will focus on incorporating the influence of corner roundedness in square capillaries to include the prediction of depinning and the associated phases which were described by Chauvet et al. \cite{chauvet_depinning_2010}. Furthermore, the inclusion of buoyancy-driven convection, could enhance the model's accuracy and applicability to more complex scenarios. Experimental validation for a broader range of geometries will also be pursued to further refine the proposed method. This novel approach provides a valuable tool to understand and predict evaporation processes, contributing to improved product design and corrosion mitigation in industrial applications.
    
    % \appendix
    
    % \section{Sample Appendix Section}
    % \label{sec:sample:appendix}
    \appendix      
\section{Surface Evolver parameter settings}
    \label{appSec:artificialDepinning}
\begin{figure}[tb]
        \centering
        \includegraphics[height=0.42\linewidth]{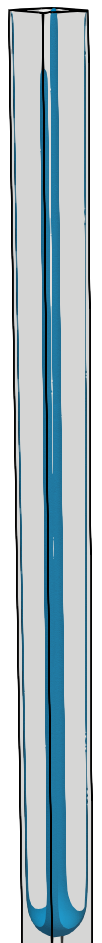}
        \includegraphics[height=0.42\linewidth]{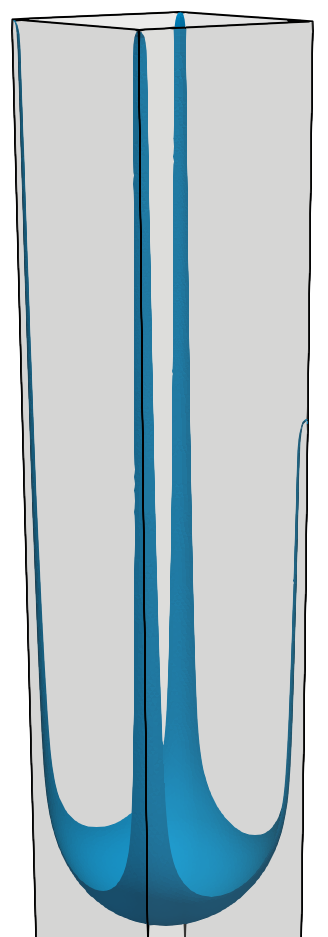}
        \includegraphics[height=0.42\linewidth]{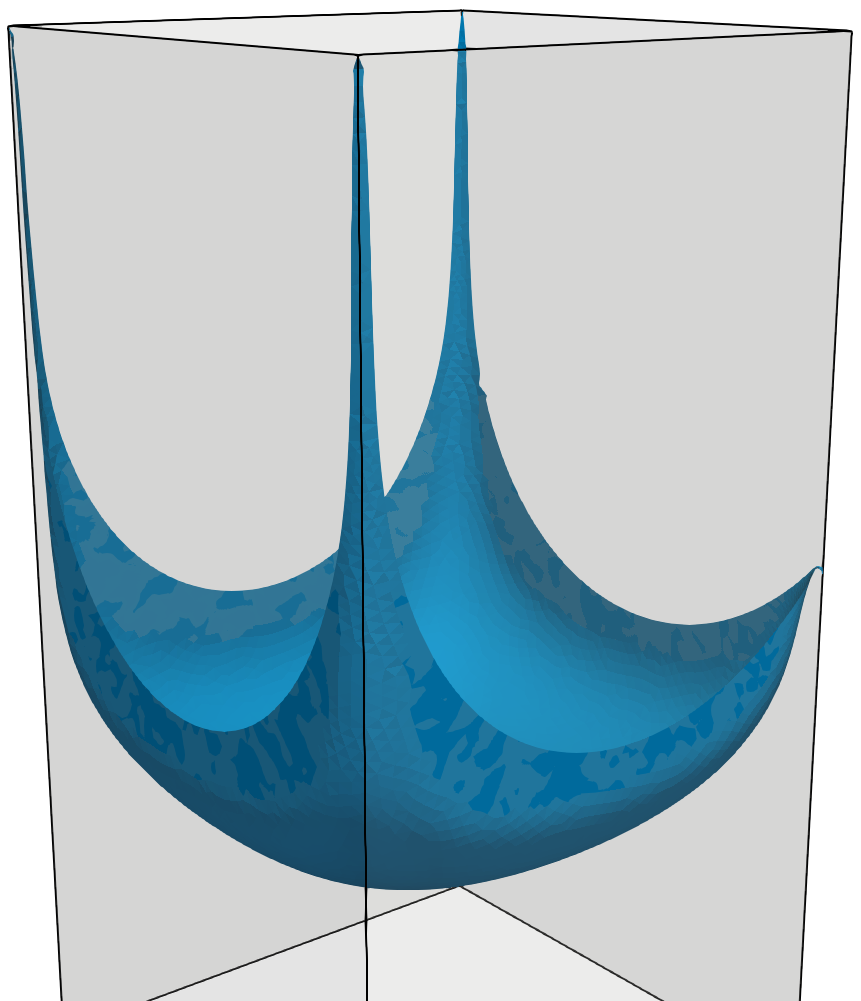}
        \includegraphics[height=0.42\linewidth]{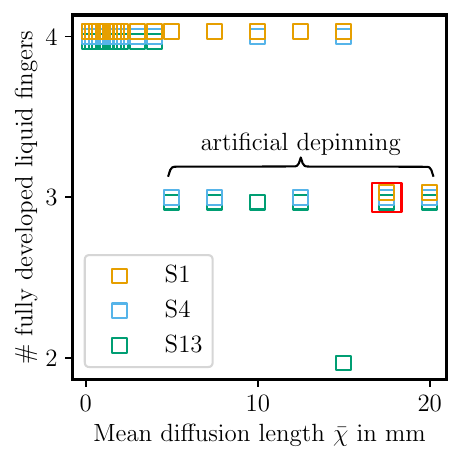}
        \caption{Artificial depinning in the cases S1, S4 and S13 (from left to right) for a mean diffusion length of $\bar{\chi}=17.5\unit{mm}$. The surfaces are simulated for a capillary height of $h\ped{c}=50\unit{mm}$ and a contact angle of $\theta=20^{\circ}$. The plot on the right shows the number of liquid fingers that rise to the top of the capillary in the cases S1, S4, S13 for various mean diffusion lengths. The surfaces on the left are marked in the plot on the right.}
        \label{appFig:artificialDepinning}
    \end{figure}

    As described in \cref{sec:methodology}, the SE simulates the liquid surface based on a gradient descent approach. Therefore, a velocity vector $\bm{v}$ is calculated for each vertex, based on the contribution of a vertex to the minimization of the total surface energy \cite{brakke_surface_2013}. In each iteration a vertex is moved along the velocity vector according to
    ~
    \begin{equation}
        \label{appEq:vertexMovement}
        \bm{x}_{i+1} = \bm{x}_i+\bm{v}s
    \end{equation}
    
    where $\bm{x}$ is the vertex position and $s$ is a universal scale factor that can be interpreted as a learning rate or time step for the algorithm \cite{brakke_surface_2013}. In order to increase convergence speed, the scale factor is optimized each iteration, whereas $s$ is limited to $s\ped{max}=1$ by default. Although this setting yields good results for low mean diffusion lengths, artificial depinning of liquid fingers in the corners of square capillaries can be observed for high mean diffusion lengths, as shown in \cref{appFig:artificialDepinning}.
    
    In the investigated cases, the artificial depinning is only observed for high mean diffusion lengths and tends to occur more often in cases where the liquid finger thickness is small in relation to the capillary cross-sectional area. For the investigated cases and mean diffusion lengths the number of liquid fingers that rise to the top of the capillary is plotted in \cref{appFig:artificialDepinning} on the right side. In the S1 case, only the simulations for $\bar{\chi}>15\unit{mm}$ have less than 4 fingers whereas the artificial depinning occurs irregularly for $\bar{\chi}>4\unit{mm}$ in the S4 and S13 case. According to the diagram in \cref{appFig:artificialDepinning}, the investigated simulations show no clear dependency on the mean diffusion length at which all liquid fingers are no longer fully developed.

    To calculate the velocity vector $\bm{v}$ of a vertex, the surface energy gradient is multiplied with the surface tension. The surface tension of water is in the order of $10^{-2}\unit{N/m}$ which results in a velocity vector of similar order of magnitude. This leads to the vertices moving too little to reach a global energy minimum state, thus a local minimum is found which incorporates depinned liquid fingers. This behavior can be prevented by using a scale limit of $s\ped{max}=1/\sigma$ \cite{brakke_surface_2013}. By scaling the velocity vector with the inverse of the surface tension, the velocity vector is guaranteed to be in the order of $1$ and allows the gradient descent approach to find the global minimum rather than a local minimum. A scale limit of $s\ped{max}=1/\sigma$ is used throughout this work which solves the problem of artificial depinning.

\section{Reliability}
    \label{appSec:reliability}
    The method presented in this paper allows the user to simulate evaporation from small cavities in a very fast way. However, it also has some limitations. SE has a particularly difficult time finding a steady-state solution in rare cases (see \cref{sec:runtime}), if the initial surface is close to the top of the capillary.

\begin{figure}[tb]
        \centering
        \begin{subfigure}[b]{0.36\linewidth}
            \centering
            \fontsize{9pt}{10pt}\selectfont
            \def\svgwidth{0.8\linewidth}
            %% Creator: Inkscape 1.3.2 (091e20e, 2023-11-25, custom), www.inkscape.org
%% Accompanies image file 'Figure_B16a.pdf' (pdf, eps, ps)
\begingroup%
  \makeatletter%
  \providecommand\color[2][]{%
    \errmessage{(Inkscape) Color is used for the text in Inkscape, but the package 'color.sty' is not loaded}%
    \renewcommand\color[2][]{}%
  }%
  \providecommand\transparent[1]{%
    \errmessage{(Inkscape) Transparency is used (non-zero) for the text in Inkscape, but the package 'transparent.sty' is not loaded}%
    \renewcommand\transparent[1]{}%
  }%
  \providecommand\rotatebox[2]{#2}%
  \newcommand*\fsize{\dimexpr\f@size pt\relax}%
  \newcommand*\lineheight[1]{\fontsize{\fsize}{#1\fsize}\selectfont}%
  \ifx\svgwidth\undefined%
    \setlength{\unitlength}{209.24774651bp}%
    \ifx\svgscale\undefined%
      \relax%
    \else%
      \setlength{\unitlength}{\unitlength * \real{\svgscale}}%
    \fi%
  \else%
    \setlength{\unitlength}{\svgwidth}%
  \fi%
  \global\let\svgwidth\undefined%
  \global\let\svgscale\undefined%
  \makeatother%
  \begin{picture}(1,1.41353539)%
    \lineheight{1}%
    \setlength\tabcolsep{0pt}%
    \put(0,0){\includegraphics[width=\unitlength,page=1]{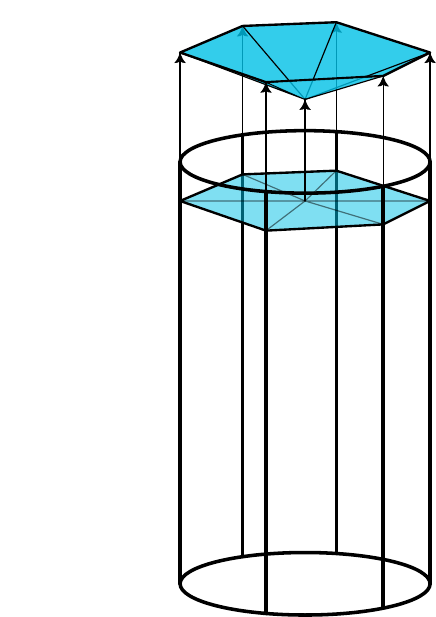}}%
    \put(0.36649107,0.80916962){\color[rgb]{0,0,0}\makebox(0,0)[rt]{\lineheight{1.25}\smash{\begin{tabular}[t]{r}initial surface\end{tabular}}}}%
    \put(0,0){\includegraphics[width=\unitlength,page=2]{Figure_B16a.pdf}}%
    \put(0.17966232,1.36968414){\color[rgb]{0,0,0}\makebox(0,0)[t]{\lineheight{1.25}\smash{\begin{tabular}[t]{c}surface after\\first iteration\end{tabular}}}}%
    \put(0.39778685,1.09873161){\color[rgb]{0,0,0}\makebox(0,0)[rt]{\lineheight{1.25}\smash{\begin{tabular}[t]{r}$\bm{v}_is$\end{tabular}}}}%
    \put(0,0){\includegraphics[width=\unitlength,page=3]{Figure_B16a.pdf}}%
  \end{picture}%
\endgroup%

            \caption{Initialization issues without upper capillary limit and initial refinement.}
            \label{appFig:firstIterationMovement} 
        \end{subfigure}
        \hfill
        \begin{subfigure}[b]{0.19\linewidth}
            \centering
            \includegraphics[width=0.16\linewidth]{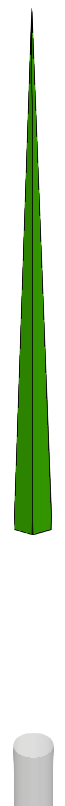}
            \includegraphics[width=0.16\linewidth]{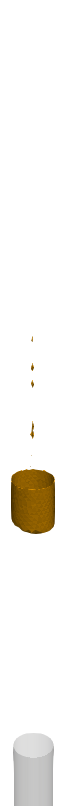}
            \caption{Without upper capillary limit and without initial refinement.}
            \label{appFig:R1_g_noUpLim}
        \end{subfigure}
        \hfill
        \begin{subfigure}[b]{0.17\linewidth}
            \centering
            \includegraphics[width=0.4\linewidth]{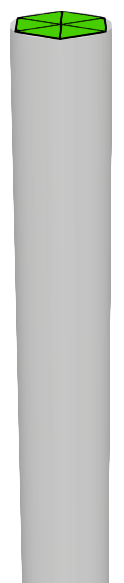}
            \includegraphics[width=0.4\linewidth]{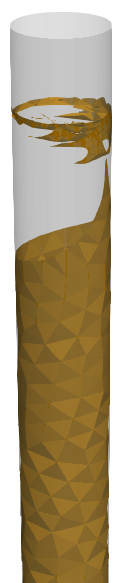}
            \caption{With upper capillary limit and without initial refinement.}
            \label{appFig:R1_g}
        \end{subfigure}
        \hfill
        \begin{subfigure}[b]{0.18\linewidth}
            \centering
            \includegraphics[width=\linewidth]{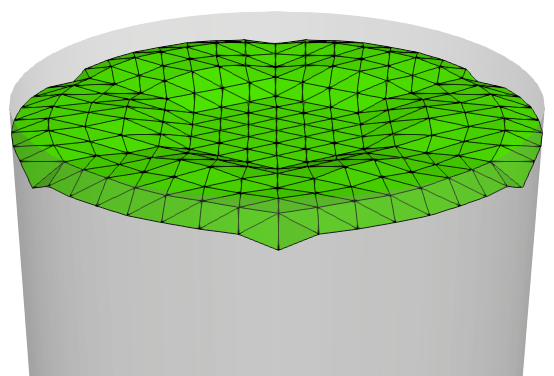}
            \includegraphics[width=\linewidth]{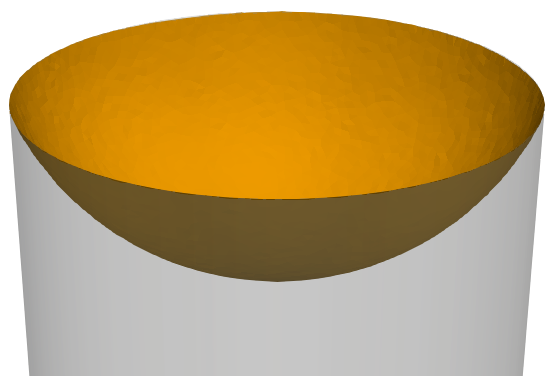}
            \caption{With upper capillary limit and with initial refinement.}
            \label{appFig:R1_rrrg}
        \end{subfigure}
        \caption{Effects of different initialization methods on the resulting liquid surface. In (b)-(d), the green surface with edges shows the solution after the first iteration and the orange surface shows the solution after running the SE-FIT optimization algorithm.}
        \label{appFig:InitializationIssues}   
    \end{figure}

    Usually the initialized water surface does not meet the contact angle boundary condition, thus the initial water surface will be far away from a fully developed state. This results in high surface energy gradients $\nabla E\ped{tot}$ in the first iterations that further lead to large vertex movement $\bm{v}_is$, since $\bm{v}_i\propto\nabla E\ped{tot}$ in \cref{appEq:vertexMovement}. Large vertex movements are particularly detrimental if the initialized surface is close to the top of the capillary. This leads to the velocity vector moving vertices outside the prescribed geometry, as shown in \cref{appFig:firstIterationMovement}, which triggers instabilities and results in unreasonable surfaces (see \cref{appFig:R1_g_noUpLim,appFig:R1_g}).

    In \cref{appFig:InitializationIssues}, the surfaces are initialized at $\bar{\chi}=0.2\unit{mm}$ followed by a different amount of refinement and iteration steps. \cref{appFig:R1_g_noUpLim} shows the result for a capillary without an explicit upper boundary condition $z_i<h\ped{c}$ where $z_i$ is the vertices z coordinate. After the first iteration step, the surface already diverges and breaks out of the capillary. Further iterations lead to a stable solution above the capillary although the volume conservation is violated. Numerical errors result in liquid droplets above the actual liquid surface.
    
    \cref{appFig:R1_g} shows the resulting surface with an explicitly implemented upper surface boundary condition $z_i<h\ped{c}$. In the first iteration the vertices are projected to the upper boundary, demonstrating the conditions effectiveness. However, the convergence algorithm leads to a very distorted and unreasonable surface.

    To obtain reasonable results, the relation $\bm{v}_i\propto l\ped{e,min,i}$, where $l\ped{e,min,i}$ is the minimum edge length at a vertex, is used to reduce the magnitude of the velocity vector at the start of a simulation. Therefore, each initial surface is refined, followed by SE iteration steps before starting the actual SE-FIT convergence algorithm. Three refinement steps, followed by 100 SE iterations showed the best results, as visualized in \cref{appFig:R1_rrrg}, and is therefore used throughout this work.

\color{\revonecolor}
\section{Surface Evolver Iteration Process}
    \label{appSec:SEIterationProcess}
\begin{figure}[tb]
        \centering
        \fontsize{11pt}{11pt}\selectfont
        \def\svgwidth{\linewidth}
        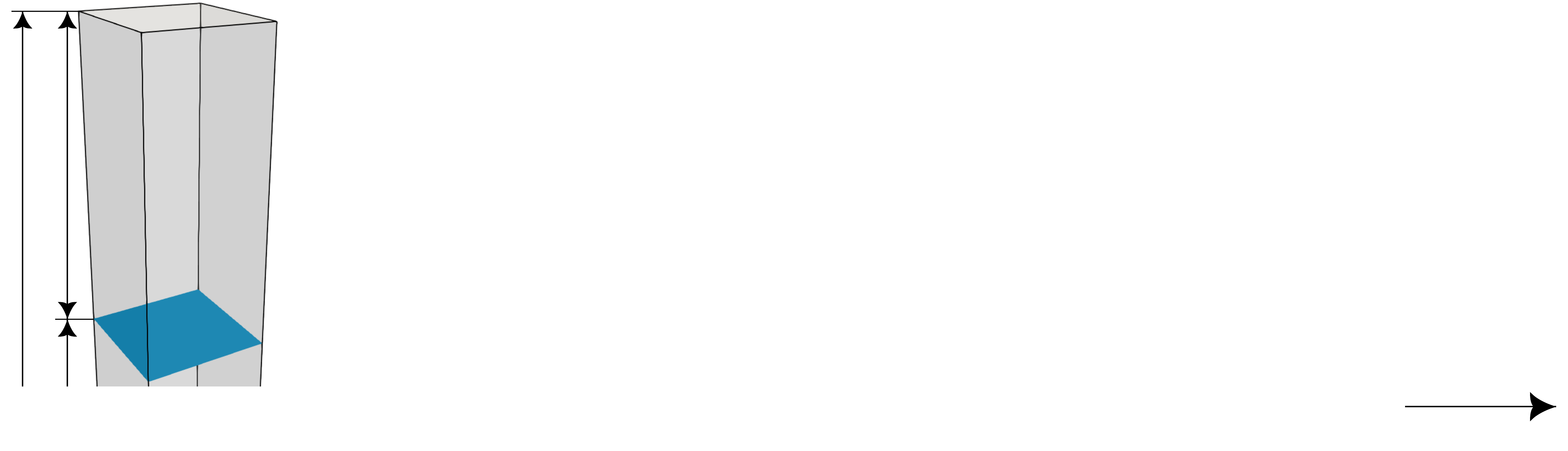
        \caption{Liquid surface shape in a square capillary of width $d\ped{c}=4\unit{mm}$ at different stages of iterations during the surface formation simulation in Surface Evolver. (left) Initialized surface at $\bar{\chi}=10\unit{mm}$ or $h\ped{l}=40\unit{mm}$ respectively, (right) fully developed liquid surface after termination criteria in SE-FIT algorithm is reached.}
        \label{appFig:SEIterationProcess}
    \end{figure}

    For the surface formation simulations the liquid surface is initialized as a plane at the height $z=h\ped{l}=h\ped{c}-\bar{\chi}$. The initial surface for a square capillary with a width of $d\ped{c}=4\unit{mm}$ at $\bar{\chi}=10\unit{mm}$ or $h\ped{l}=40\unit{mm}$ respectively, is illustrated on the left of \cref{appFig:SEIterationProcess}. Further, the figure shows the evolving surface every 50 iterations and the steady state surface on the right side.
    
    As shown in \cref{appFig:SEIterationProcess}, the liquid fingers in the corner of the capillary evolve rapidly in the beginning until they reach the upper capillary limit at $z=h\ped{c}$. After 250 iterations, the liquid surface only changes slightly to achieve a steady state solution. The termination criteria for a steady state solution is defined as

    \begin{align}
        \label{appEq:meanSurfaceEnergy}
        \bar{E_{\gamma}}^{i} = \frac{1}{5}\sum_{j=i-4}^i E_{\gamma}^j,\\
        \label{appEq:SETerminationCriteria}
        \bar{E_{\gamma}}^{i+1} - \bar{E_{\gamma}}^{i} < 10^{-6}.
    \end{align}

    where $\bar{E_{\gamma}}^{i}$ is the mean surface energy at iteration $i$ over the last 5 iterations and $E_{\gamma}^j$ is the surface energy in iteration $j$.
\color{black}

    \section*{Declaration of generative AI and AI-assisted technologies in the writing process}

During the preparation of this work, the authors used ChatGPT and DeepL in order to improve the readability and perform grammar checking. After using this tool/service, the authors reviewed and edited the content as needed and take full responsibility for the content of the published article.

    \section*{Acknowledgements}
We gratefully thank Tobias Tolle for his invaluable guidance and support throughout this project. His expertise in coding and insightful supervision were instrumental in the development and implementation of the simulation methods. We especially appreciate his willingness to address our coding questions and provide helpful direction.
    
     \bibliographystyle{elsarticle-num} 
     \bibliography{Paper1-LitLibrary}
    
    
    
    
    
    % \end{thebibliography}
\end{document}